\documentclass[12pt]{article}
\pdfoutput=1

\usepackage{hyperref}
\hypersetup{colorlinks=true,
  pdfborder={0 0 0},
  linktocpage=false
}

\usepackage[normalem]{ulem}
\usepackage[utf8]{inputenc}
\usepackage[left=2.55cm, right=2.55cm, top=2.55cm, bottom=2.55cm]{geometry}
\usepackage{amsmath,amssymb}
\usepackage{slashed}
\usepackage{xcolor}
\usepackage{graphicx}
\usepackage{url}
\usepackage{cancel}
\usepackage{cite}
\usepackage{tabularx,booktabs}
\usepackage{multicol}
\usepackage{units}
\usepackage{xspace}
\usepackage[labelfont=bf]{caption}

\usepackage{array,multirow}
\usepackage{booktabs}
\usepackage{colortbl}
\usepackage{float}

\definecolor{darkred}{rgb}{0.6,0,0}
\definecolor{darkpurple}{rgb}{0.5,0,0.5}
\newcommand {\red} {\color{red}}

\def\gsim{\raise0.3ex\hbox{$\;>$\kern-0.75em\raise-1.1ex\hbox{$\sim\;$}}}
\def\lsim{\raise0.3ex\hbox{$\;<$\kern-0.75em\raise-1.1ex\hbox{$\sim\;$}}}

\newcommand{\crowcolor}{\rowcolor[rgb]{0.9,0.9,0.9}}
\newcommand{\ctoprule}{\toprule[0.5mm]}
\newcommand{\cbottomrule}{\bottomrule[0.5mm]}

\begin{document}

%\vspace*{-2cm}
%\begin{flushright}
%IFIC/20-XX \\
%\vspace*{2mm}
%\today
%\end{flushright}

\begin{center}
\vspace*{0mm}

\vspace{-1.cm}
{\Large \bf 
How many 1-loop neutrino mass models are there?} \\
\vspace{3mm}

{\bf Carolina Arbel\'aez$^{\text{a}}$, Ricardo Cepedello$^{\text{b}}$,
Juan Carlos Helo$^{\text{c}, \text{d}}$, Martin Hirsch$^{\text{e}}$, Sergey Kovalenko$^{\text{d}, \text{f}}$  }

\vspace*{.5cm} $^{\text{a}}$ Department of Physics, Universidad
T\'ecnica Federico Santa Mar\'ia and Centro Cient\'ifico Tecnol\'ogico
de Valpara\'iso CCTVal, \protect\\ Avenida Espa\~na 1680,
Valpara\'iso, Chile

\vspace* {.5cm}{$^{\text{b}}$ Institut f\"ur Theoretische Physik und 
Astrophysik, \\
University of W\"urzburg,  Campus Hubland Nord, D-97074 W\"urzburg, 
Germany 

\vspace*{.5cm} $^{\text{c}}$ Departamento de F\'isica 
, Facultad de Ciencias, \\ Universidad de la Serena,
Avenida Cisternas 1200, La Serena, Chile

\vspace*{.5cm} $^{\text{d}}$ Millennium Institute for Subatomic Physics at the High Energy Frontier (SAPHIR), Fern\'andez
Concha 700, Santiago, Chile

\vspace*{.5cm} $^{\text{e}}$Instituto de F\'{\i}sica Corpuscular
(CSIC-Universitat de Val\`{e}ncia), \\ C/ Catedr\'atico Jos\'e
Beltr\'an 2, E-46980 Paterna (Val\`{e}ncia), Spain

\vspace*{.5cm} $^{\text{f}}$ Departamento de Ciencias F\'isicas, Universidad Andr\'es Bello, Sazie 2212, Piso 7, Santiago, Chile

 \vspace*{.3cm}
 \href{mailto:carolina.arbelaez@usm.cl}{carolina.arbelaez@usm.cl},
 \href{mailto:ricardo.cepedello@physik.uni-wuerzburg.de}{ricardo.cepedello@physik.uni-wuerzburg.de}},
 \href{mailto:jchelo@userena.cl}{jchelo@userena.cl},
 \href{mailto:mahirsch@ific.uv.es}{mahirsch@ific.uv.es},
  \href{mailto:sergey.kovalenko@unab.cl}{sergey.kovalenko@unab.cl}

\end{center}

\vspace*{3mm}
\begin{abstract}\noindent\normalsize
It is well-known that at tree-level the d=5 Weinberg operator can be
generated in exactly three different ways, the famous seesaw
models. In this paper we study the related question of how many
phenomenologically consistent 1-loop models one can construct at
d=5. First, we discuss that there are two possible classes of 1-loop
neutrino mass models, that allow avoiding stable charged relics:
(i) Models with dark matter candidates and (ii) models with
``exits''. Here, we define ``exits'' as particles that can decay into
standard model fields. Considering 1-loop models with new scalars and
fermions, we find in the dark matter class a total of (115+203) models,
while in the exit class we find (38+368) models.  Here, 115 is the
number of DM models, which require a stabilizing symmetry, while 203
is the number of models which contain a dark matter candidate, which
maybe accidentally stable. In the exit class the 38 refers to models, for
which one (or two) of the internal particles in the loop is a SM
field, while the 368 models contain only fields beyond the SM (BSM) in
the neutrino mass diagram. We then study the RGE evolution of the
gauge couplings in all our 1-loop models. Many of the models in our
list lead to Landau poles in some gauge coupling at rather low
energies and there is exactly one model which unifies
the gauge couplings at energies above $10^{15}$ GeV in a numerically 
acceptable way.

\end{abstract}

%\newpage

%\tableofcontents

%\newpage

\section{Introduction\label{sect:intro}}

It is well-known that the Weinberg operator, ${\cal O}_W$,
\cite{Weinberg:1979sa} can be generated at tree-level in exactly three
different ways \cite{Ma:1998dn}, the famous seesaw mechanisms
\cite{Minkowski:1977sc,Yanagida:1979as,Mohapatra:1979ia,GellMann:1980vs,
  Schechter:1980gr,Mohapatra:1980yp,Foot:1988aq}.  The smallness of
the observed neutrino masses has also motivated many papers on
radiative neutrino mass models, starting with the classical papers
\cite{Zee:1980ai,Cheng:1980qt,Zee:1985rj,Babu:1988ki}.  For a recent
review on loop models for neutrino mass, see \cite{Cai:2017jrq}. One
interesting question to ask then naturally is: How many possibilities
actually exist to generate ${\cal O}_W$ at 1-loop level? We will
explore this question in the current paper. 

\begin{figure}[t]
\centering
\includegraphics[width=0.49\textwidth]{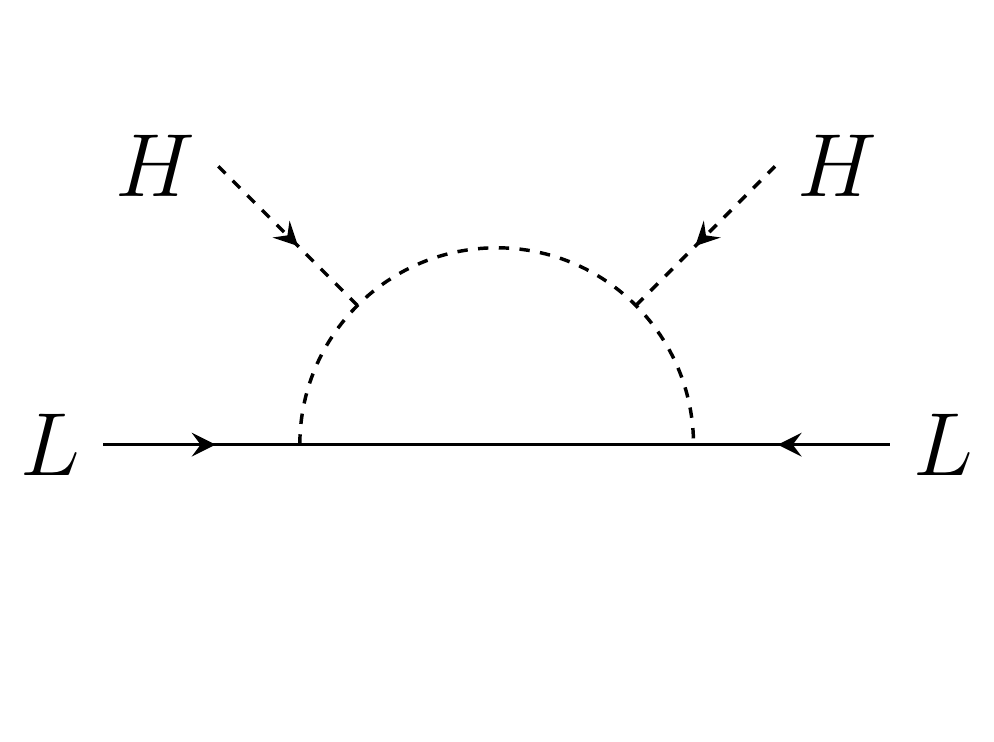}
\includegraphics[width=0.49\textwidth]{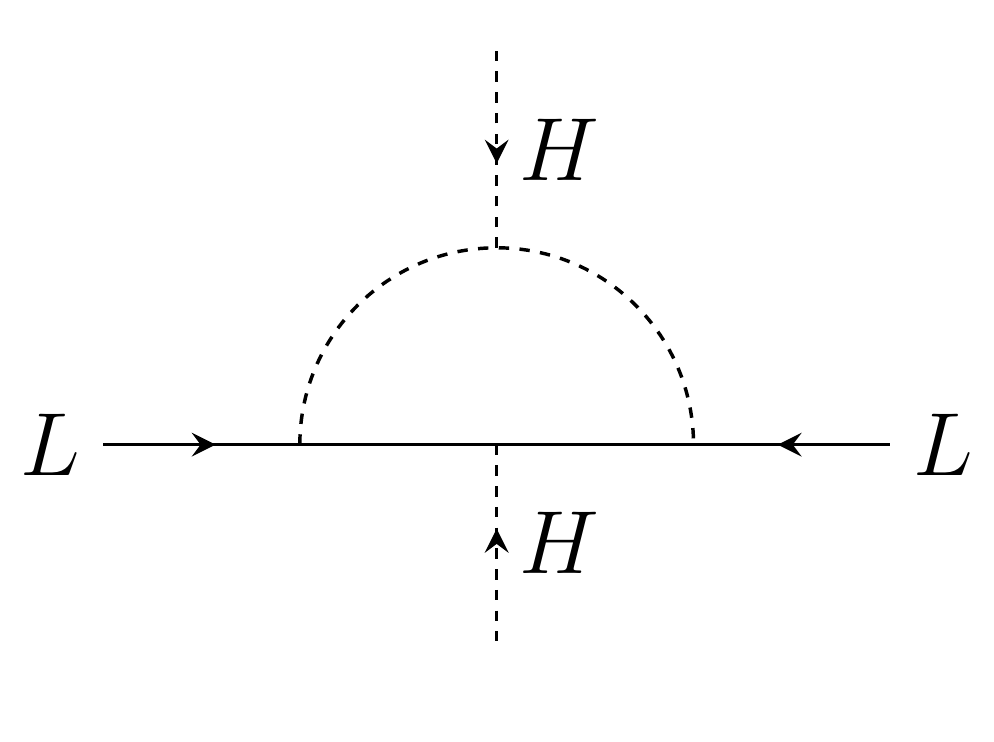}
\\
\vskip3mm
\includegraphics[width=0.49\textwidth]{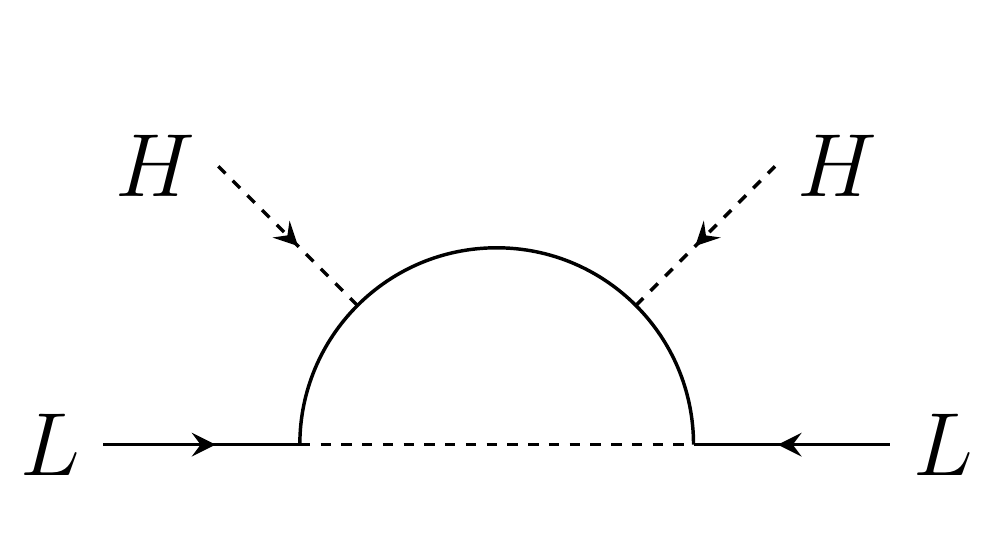}
\includegraphics[width=0.49\textwidth]{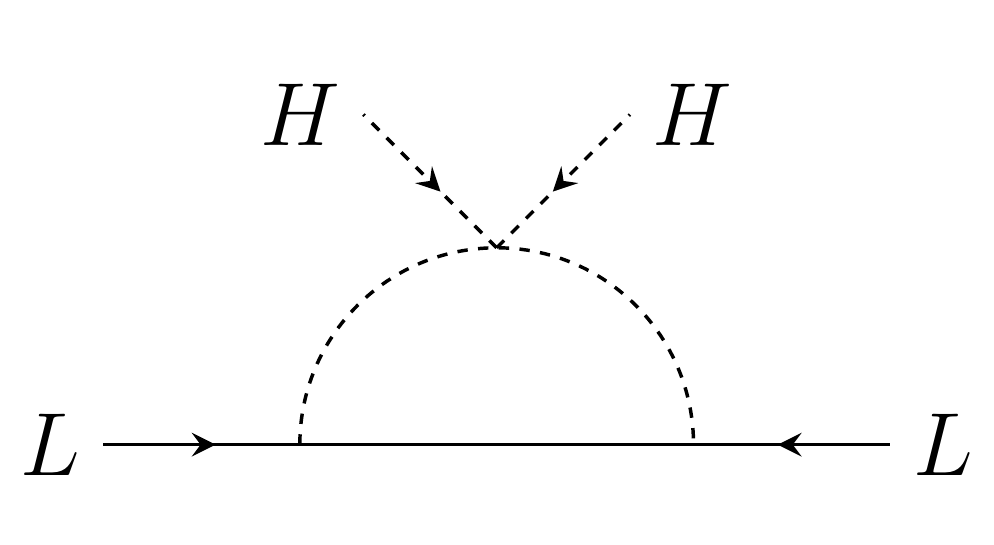}
\caption{The four different 1-loop diagrams that can lead to genuine
  neutrino mass models \cite{Bonnet:2012kz}. Top line: T-I-1 (left)
  and T-I-2 (right), bottom T-I-3 (left) and T-3 (right).}
\label{fig:GenDiag}
\end{figure}

Partial answers to our question already exist in the literature,
starting with \cite{Ma:1998dn}. Ref. \cite{Bonnet:2012kz} worked
within a diagrammatic approach: Construct all possible topologies and
from there derive all possible diagrams that can lead to ${\cal O}_W$.
Ref. \cite{Bonnet:2012kz} found a total of only four diagrams,
descending from two topologies, which can yield ``genuine'' neutrino
mass models, see figure \ref{fig:GenDiag}.
\footnote{A similar approach was followed in
  \cite{Sierra:2014rxa,Cepedello:2019zqf} for 2-loop models and in
  \cite{Cepedello:2018rfh} for 3-loop models.}  Here, ``genuine''
models are defined as models for which at n-loop level all
contributions to ${\cal O}_W$ with n-1 loops or less (can) vanish,
i.e. the list of diagrams in these constructions do not contain
self-energies and other loop diagrams that are guaranteed to be only
corrections or subdominant to lower order contributions. Using these
criteria ref. \cite{Bonnet:2012kz} then provides lists of all 1-loop
models up to electro-weak triplets. Note also that \cite{Restrepo:2013aga} 
lists 1-loop neutrino mass models with dark matter candidates 
up to electro-weak triplets.

An alternative approach to the problem is based on effective field 
theory. Here, one first constructs all lepton number violating
operators, starting with ${\cal O}_W$ and up to the desired dimension,
allowed by the SM field content and symmetries.
Ref. \cite{Babu:2001ex} lists all $\Delta(L)=2$ operators up to
dimension $d=11$, see also \cite{deGouvea:2007qla}. ``Opening up'' or
``exploding'' the operators in all possible ways, together with adding
the appropriate numbers of Higgses to close loops, then, in principle, 
also allows a systematic construction of neutrino masses models
\cite{Gargalionis:2020xvt}.

None of the above papers, however, gives a ``complete'' list of 1-loop
models. While ref. \cite{Bonnet:2012kz} provided a partial list of
1-loop neutrino mass models, here we aim at giving the complete list
of {\em phenomenologically consistent} models. Of course, all models
for neutrino mass should be able to reproduce experimental neutrino
data, see for example \cite{deSalas:2020pgw}, while at the same time
obey upper limits on charged lepton flavour violation searches
\cite{ParticleDataGroup:2020ssz}. However, there are also other
criteria that a successful model should fulfill and our main concern
here is to avoid problems with cosmology. 

Consider the following, very basic observation: In loop models of
${\cal O}_W$ the particles internal to the n-loop diagram always
couple in pairs to some external SM field, compare
figure~\ref{fig:GenDiag}. If all particles in the loop are fields
beyond the SM (BSM) and there are no other interactions in the model
for the BSM fields than those appearing in the loop diagram, the
corresponding model will have some accidental symmetry (in the
simplest case a $Z_2$). The lightest of the loop particles (LLP) will
then be absolutely stable.

Stable BSM fields, however, might lead to drastic changes in
cosmology, in particular if they are electrically charged (and/or
coloured).  Experimental searches for stable charged relics put severe
bounds on their abundance in the mass range $M \sim [1, 10^5]$ GeV,
see for example
\cite{ParticleDataGroup:2020ssz,Hemmick:1989ns,Kudo:2001ie,Taoso:2007qk}.
To avoid problems with cosmology, (at least one of) the BSM particles
in 1-loop models should therefore be able to decay to SM fields --
unless the lightest of them is electrically neutral. This simple
consideration limits the number of allowed, electrically charged BSM
particles, see section~\ref{subsect:exit}, and thus one can have only
a finite set of 1-loop neutrino mass models.

Electrically neutral BSM particles, on the other hand, can be
candidates for dark matter in the form of WIMPs. Thus, one can
construct 1-loop models of neutrino mass in which the lightest loop
particles is a WIMP candidate, instead of decaying to SM fields. One
can write down an infinite number of electro-weak multiplets, that
contain one neutral state. However, the list of phenomenologically
acceptable multiplets for WIMP candidates is rather short. To the best
of our knowledge, this was first discussed in \cite{Cirelli:2005uq}
and the subject has very recently been reconsidered in
\cite{Bottaro:2021snn,Bottaro:2022one}, see also \cite{Hambye:2009pw}
and \cite{Belyaev:2022qnf}. Note that the recent papers
\cite{Bottaro:2021snn, Bottaro:2022one} allows a larger list of
acceptable multiplets than \cite{Cirelli:2005uq}.
The possible connection between dark matter in $SU(2)$ representations
larger than triplets and loop models of neutrino mass has been
discussed before in
\cite{Cai:2011qr,Kumericki:2012bf,Cai:2016jrl,Ahriche:2016rgf,AristizabalSierra:2016vog}. We will discuss more details in section \ref{subsect:dm}.

We thus will discuss two types of loop models not disfavoured by
cosmology:\footnote{Somewhat fuzzily we call these
  ``phenomenologically consistent'' models, see discussion in
  section~\ref{sect:sum}.} (i) ``exit'' models, i.e. models in which
there are no stable particles in the loop and (ii) dark matter models,
i.e. models in which one of the loop particles can be a good WIMP
candidate. For both cases we construct the complete list of models. We
find that there are (38+368) exit models, while in the dark matter
class there are a total of (115+203) models.  In the exit class the 38
refers to models, for which one (or more) of the internal particles in
the loop is a SM field, while the 368 models contain only fields
beyond the SM. The 115 is the number of DM models, which require an
additional symmetry to give an acceptable WIMP candidate, while 203 is
the number of models which contain a dark matter candidate, which maybe 
accidentally stable.  \\

The rest of the paper is organized as follows. In the next section we
discuss the concrete criteria applied in the construction of our
models. Subsection~\ref{subsect:exit} deals with models in the exit
class, while \ref{subsect:dm} discusses the dark matter class. In
section~\ref{sect:rge} we then turn to RGE evolution of the SM gauge 
couplings. Since many of our models contain large $SU(2)$ and/or
coloured multiplets, Landau poles at rather low energies appear in
many of these constructions. Interestingly in our long list of models
there is exactly one variant, which leads to a near-perfect
unification of the gauge couplings at a scale of roughly $m_G \simeq 10^{17}$ 
GeV. In section~\ref{sect:sum}, we summarise briefly
our results and discuss how our lists would change, modifying or
dropping some of the assumptions that went into their
construction. The complete lists of models are relegated to the
appendix.

\section{Setup and models\label{sect:setup}}

Following \cite{Bonnet:2012kz} in our discussion we will concentrate
on models with scalars and fermions. For models with new gauge
vectors, the standard model gauge symmetry has to be extended and that
symmetry needs to be broken to the SM. This implies that the scalar
sector of the model needs to be discussed as well. This is beyond the
scope of our present work. We will, however, briefly discuss loops
with vectors in section~\ref{sect:sum}.  Note that models with new
vectors will not require any additional diagram, beyond those shown in
figure~\ref{fig:GenDiag}.  This section is divided into two parts. We
will examine first models with exits, before turning to dark matter
models.

\subsection{Exit models\label{subsect:exit}}

In this subsection we will discuss the construction of models that
contain no new stable particle, i.e. ``exit'' models.  In this case,
all particles in the neutrino mass loop can be charged and/or
coloured. No new symmetry, beyond those of the SM, is needed to make
these models genuine, if we also demand that the particle content of a
given model does not generate $O_W$ at tree-level. Thus, from our list
of models in this class we delete all possible constructions
containing either $F_{1,1,0}=N_R$, $F_{1,3,0}=\Sigma$ or
$S_{1,3,1}=\Delta$.\footnote{One can avoid the tree-level generation
  of $O_W$ also using an additional (discrete) symmetry.  Models with
  additional symmetries can contain $N_R$, $\Sigma$ and $\Delta$, they
  are discussed in the next subsection. Here, we only mention that one
  can construct, in principle, an additional 78 1-loop models in the
  exit class using these fields. We disregard all of them in the
  following as ``non-genuine.''} Here and elsewhere in this paper, we
use the notation $F$ or $S$ to denote fermions or scalars, with the
subscript showing the transformation properties and quantum numbers
for the SM gauge group, $SU(3)_C\times SU(2)_L \times
U(1)_Y$. Alternatively, for compactness, we use the notation
introduced in \cite{deBlas:2017xtg}, see also tables~\ref{t:scalars}
and \ref{t:fermions}.

%%%%%%%%%%%%%%%%%%%%%%%%%%%%%%
% TABLE SCALARS

%\setlength{\aboverulesep}{0pt}
%\setlength{\belowrulesep}{0pt}

\begin{table}[t]
  \begin{center}
    {\small
      \begin{tabular}{lcccccccc}
        \ctoprule
        \crowcolor
        Name &
        ${\cal S}^{(a)}$ &
        ${\cal S}_1$ &
        ${\cal S}_2$ &
        $\varphi$ &
        $\Xi$ &
        $\Xi_1/\Delta^{(a,b)}$ &
        $\Theta_1^{(c)}$ &
        $\Theta_3^{(c)}$ \\
        Irrep &
        $\left(1,1,0\right)$ &
        $\left(1,1,1\right)$ &
        $\left(1,1,2\right)$ &
        $\left(1,2,{\frac 12}\right)$ &
        $\left(1,3,0\right)$ &
        $\left(1,3,1\right)$ &
        $\left(1,4,{\frac 12}\right)$ &
        $\left(1,4,{\frac 32}\right)$ \\[1.3mm]
        \cbottomrule
        &&&&&&&\\[-0.4cm]
        \ctoprule
        \crowcolor
        Name &
        ${\omega}_{1}$ &
        ${\omega}_{2}$ &
        ${\omega}_{4}$ &
        $\Pi_1$ &
        $\Pi_7$ &
        $\zeta$ &
        & \\
        Irrep &
        $\left(3,1,{-\frac 13}\right)$ &
        $\left(3,1,{\frac 23}\right)$ &
        $\left(3,1,{-\frac 43}\right)$ &
        $\left(3,2,{\frac 16}\right)$ &
        $\left(3,2,{\frac 76}\right)$ &
        $\left(3,3,{-\frac 13}\right)$ \\[1.3mm]
        \cbottomrule
        &&&&&&&\\[-0.4cm]
        \ctoprule
        \crowcolor
        Name &
        $\Omega_{1}$ &
        $\Omega_{2}$ &
        $\Omega_{4}$ &
        $\Upsilon$ &
        $\Phi$ &
        &
        & \\
        Irrep &
        $\left(6,1,{\frac 13}\right)$ &
        $\left(6,1,{-\frac 23}\right)$ &
        $\left(6,1,{\frac 43}\right)$ &
        $\left(6,3,{\frac 13}\right)$ &
        $\left(8,2,{\frac 12}\right)$ \\[1.3mm]
        \cbottomrule
      \end{tabular}
    }
    \caption{Scalar ``exits'': Scalar bosons that can couple to a pair
      of standard model fields (in case of $^{(c)}$: three Higgses).
      $^{(a)}$: the field does not appear in the list of valid 1-loop
      decompositions of ${\cal O}_W$; $^{(b)}$:
      Ref. \cite{deBlas:2017xtg} uses the symbol $\Xi_1$. In neutrino
      physics this field is usually denoted as $\Delta$ (seesaw
      type-II).}
    \label{t:scalars}
  \end{center}
\end{table}

We can divide the exit class of models into two sub-classes: (i) The
model contains at least one SM field in the loop. In this case the
model by construction does not contain any stable BSM particles.  And
(ii) all particles appearing in the loop are BSM fields. In that case,
there must be at least one particle among the BSM fields, which can
decay to SM fields.  A list of all BSM scalars that can decay to SM
fields at tree-level are given in table~\ref{t:scalars}. This table
coincides with table 1 in reference \cite{deBlas:2017xtg}.
Ref. \cite{deBlas:2017xtg} arrived at this table from a completely
different consideration, namely, from the construction of all
tree-level completions for the $d=6$ SM effective field theory
(SMEFT). The lists coincide simply because in both cases a BSM field
must appear linearly in at least one term of the Lagrangian. That term
will allow to generate a $d=6$ operator in SMEFT at tree-level and, at
the same time, is responsible for the decay of the BSM field.  In the
table we give the quantum numbers of the scalars in the order
$SU(3)_C\times SU(2)_L\times U(1)_Y$ and also the symbols proposed in
\cite{deBlas:2017xtg}.

A few comments are in order. First of all, ${\cal S}$ does not appear
in the list of genuine 1-loop exit models, we include it in the table
only for completeness.\footnote{All 1-loop diagrams with ${\cal S}$ 
will either contain $N$ or $\Xi_1/\Delta$ and thus are eliminated as 
non-genuine.} Second, the field $\Xi_1/\Delta$ does also not
appear in our list of ``genuine'' exit models, since $\Delta$ is the
mediator of the tree-level seesaw type-II.  This field has been
denoted as $\Xi_1$ in \cite{deBlas:2017xtg}, but in the neutrino
physics community it is more commonly known as $\Delta$.
\footnote{We also note that the fields ${\omega}_{1}$, ${\omega}_{4}$,
  $\Pi_1$, $\Pi_7$ and $\zeta$ are known in the literature as scalar
  ``leptoquarks''.  In the notation of the classic paper
  \cite{Buchmuller:1986zs}, these are called $S_0$, ${\tilde S}_0$,
  ${\tilde S}_{1/2}$, $S_{1/2}$ and $S_1$, respectively.} Also note
that all scalars in the table can decay to SM fermion pairs, except
${\cal S}$, ${\cal S}_1$ and $\Xi$, which decay to pair of
Higgses. Finally, $\Theta_1$ and $\Theta_3$ will decay to three
Higgses. Particularly interesting, from the point of view of model
building, is $\Theta_3$, since this quadruplet scalar appears in the
only genuine tree-level model \cite{Bonnet:2009ej,Babu:2009aq} for the
operator $O_{7,W} = (H^\dagger H)\cdot O_W$.

Table~\ref{t:fermions} contains all BSM fermion fields that can couple
to a SM fermion plus a Higgs. In the mass eigenstate basis, in
addition to decays to the physical Higgs and a SM fermion, these
fields will also decay to a SM fermion plus a gauge boson.  Again, we
include $F_{1,1,0}=N$ and $F_{1,3,0}=\Sigma$ for completeness,
although model constructions involving these fields are not included
in our lists of genuine exit models, since they generate tree-level
seesaws. There are five fields which have quantum numbers coinciding
with some SM fermion. However, all fields in table~\ref{t:fermions}
should be understood as vector-like fields (or self-conjugate Majorana
fields, in case of $N$ and $\Sigma$). We do not write the vector-like
partners explicitly.

%
%%%%%%%%%%%%%%%%%%%%%%%%%%%%%%
% TABLE FERMIONS
%
\begin{table}[t]
  \begin{center}
    {\small
      \begin{tabular}{lccccccc}
        \ctoprule
        \crowcolor
        Name &
     $N^{(a)}$ & $E$ & $\Delta_1$ & $\Delta_3$ & $\Sigma^{(a)}$ & $\Sigma_1$ & \\
        Irrep &
        $\left(1, 1,0\right)$ &
        $\left(1, 1,-1\right)$ &
        $\left(1, 2,-\frac{1}{2}\right)$ &
        $\left(1, 2,-\frac{3}{2}\right)$ &
        $\left(1, 3,0\right)$ &
        $\left(1, 3,-1\right)$ & \\[1.3mm]
       \cbottomrule
        &&&&&&&\\[-0.4cm]
        \ctoprule
        \crowcolor
        Name &
        $U$ & $D$ & $Q_1$ & $Q_5$ & $Q_7$ & $T_1$ & $T_2$ \\
        Irrep &
        $\left(3, 1,\frac{2}{3}\right)$ &
        $\left(3, 1,-\frac{1}{3}\right)$ &
        $\left(3, 2,\frac{1}{6}\right)$ &
        $\left(3, 2,-\frac{5}{6}\right)$ &
        $\left(3, 2,\frac{7}{6}\right)$ &
        $\left(3, 3,-\frac{1}{3}\right)$ &
        $\left(3, 3,\frac{2}{3}\right)$ \\[1.3mm]
        \cbottomrule
      \end{tabular}
    }
    \caption{Fermion exits: New vector-like fermions that can couple
      to standard model fields. $^{(a)}$: field does not appear in the
      list of ordinary genuine 1-loop decompositions of ${\cal O}_W$,
      since it mediates tree-level seesaw type-I/III. Symbols are again
    taken from \cite{deBlas:2017xtg}.}
    \label{t:fermions}
  \end{center}
\vspace{1cm}
\end{table}

Given these lists of all possible BSM fields, that can decay directly
to standard model particles via renormalisable interactions at
tree-level, we can construct all possible 1-loop neutrino mass exit
model variants, using the diagrams in figure~\ref{fig:GenDiag}. The
task is in principle straight-forward, albeit tedious. We use our own
code written in {\tt Mathematica} to automatise the systematic
generation of neutrino mass models. The diagrams in
figure~\ref{fig:GenDiag} can be represented as adjacency matrices by
giving numbers to all the vertices. Each entry of the matrices will
then correspond to a field in the diagram, i.e.  entry $(i,j)$ will be
the field connecting vertices $i$ and $j$. The power of this approach
is twofold: (i) the contraction of all fields along a row or column
should contain always a singlet, and (ii) with adjacency matrices one
can then use tools from graph theory, for instance, to delete
isomorphic diagrams. The external fields are already known, so once
given the quantum number for one of the fields in the loop, the rest
can be computed. This can be actually done for a general set of
quantum numbers for the starting field (seed). As the external
particles are colour blind, all the particles in the loop will have
the same $SU(3)_C$ representation as the seed, while $SU(2)_L$ and
hypercharge can be obtained by systematically solving the set of
equations for each vertex, i.e. for each row/column of the adjacency
matrix. Note that one should keep track of the several possibilities
for the products of $SU(2)$ representations, for example, a
representation $r$ times a doublet gives two possible representations
$r\pm 1$, where only those representations larger than $\mathbf{1}$
are possible.  Numbers can then be systematically given to the free
charges of the seed to get a complete list of models, to which we
apply our genuineness criteria and further classify them, as explained
in the text. Chirality is also being tracked along the fermion line to
afterwards check whether any of the internal fermions may be a SM
fermion. It is worth noticing a slight subtlety in this approach
first shown in \cite{Cepedello:2018rfh}: The antisymmetric
contractions of $SU(2)$ implies that some couplings with identical
particles vanishes exactly, for example, the coupling of two identical
$SU(2)$ doublets (like two Higgses) to a singlet. Diagrams with such
couplings should be removed.\footnote{Even if this is a local feature
  of $SU(2)$, diagrams with such non-local (effective) couplings may
  also vanish if, for example, the identical particles get a VEV,
  which is the case of the SM Higgs.}

The resulting lists are given in table~\ref{tab:SMLps} and in the
appendix. The tables in the appendix are divided first into the four
diagrams of figure~\ref{fig:GenDiag}. The models are then ordered
first with respect to the scalar exits in table~\ref{t:scalars}, then
w.r.t. fermions as in table~\ref{t:fermions} and then sub-divided
again into increasing number of exits that occur in each diagram.

We will discuss now a few, particular cases found in those tables.  A
subset of the models appearing in the diagrams T-I-2 and T-I-3 use
fermions, with quantum numbers and couplings identical to one of 
the standard model fermion. In these cases, either one or two of
the internal particles can be identified with SM quarks or
leptons. These are particularly simple models, in the sense that fewer
BSM fields are needed than in all other cases. We have identified a
total of 38 possibilities in this special sub-class and list all of
them in table~\ref{tab:SMLps}. Note that in case there are two 
SM fields in the diagram, as for example in model $\#$5, there 
are two more models (in this example $\#$4 and $\#$8) in which one 
of the two SM fields could also be a new, vector-like particles. 
These are counted in this table as extra models, since they contain 
a different number of degrees of freedom, see also section \ref{sect:rge}.

 \begin{table}[t]
  \begin{center}
    {\tiny
\begin{tabular}[t]{|c|c||c|c||c|c|}
    \hline
     \# & Fields &  \# & Fields &  \# & Fields\\ 
    \hline
1& ${\red L}F_{1,1,1}F_{1,2,3/2}S_{1,1,1}$ &
2& ${\red Le_R}F_{1,2,3/2}S_{1,1,1}$ &
3& ${\red L}F_{1,3,1}F_{1,2,3/2}S_{1,1,1}$ \\
4& ${\red L}F_{1,1,1}S_{1,1,1}S_{1,2,1/2}$ &
5& ${\red Le_R}S_{1,1,1}S_{1,2,1/2}$ &
6& ${\red L}F_{1,3,1}S_{1,1,1}S_{1,2,1/2}$ \\
7& ${\red e_R}F_{1,2,1/2}F_{1,2,3/2}S_{1,1,1}$ &
8& ${\red e_R}F_{1,2,1/2}S_{1,1,1}S_{1,2,1/2}$ &
9& ${\red Q}F_{3,1,-1/3}F_{3,2,-5/6}S_{3,1,-1/3}$ \\
10& ${\red Q}F_{3,1,-1/3}F_{3,2,-5/6}S_{3,3,-1/3}$ &
11& ${\red Qd_R}F_{3,2,-5/6}S_{3,1,-1/3}$ &
12& ${\red Qd_R}F_{3,2,-5/6}S_{3,3,-1/3}$ \\
13& ${\red Q}F_{3,3,-1/3}F_{3,2,-5/6}S_{3,1,-1/3}$ &
14& ${\red Q}F_{3,3,-1/3}F_{3,2,-5/6}S_{3,3,-1/3}$ &
15& ${\red Q}F_{3,3,-1/3}F_{3,4,-5/6}S_{3,3,-1/3}$ \\
16& ${\red Q}F_{3,1,-1/3}F_{3,3,2/3}S_{3,2,1/6}$ &
17& ${\red Qd_R}F_{3,3,2/3}S_{3,2,1/6}$ &
18& ${\red Q}F_{3,3,-1/3}F_{3,1,2/3}S_{3,2,1/6}$ \\
19& ${\red Q}F_{3,3,-1/3}F_{3,3,2/3}S_{3,2,1/6}$ &
20& ${\red Q}F_{3,3,-1/3}F_{3,3,2/3}S_{3,4,1/6}$ &
21& ${\red Q}F_{3,1,-1/3}S_{3,2,1/6}S_{3,1,-1/3}$ \\
22& ${\red Q}F_{3,1,-1/3}S_{3,2,1/6}S_{3,3,-1/3}$ &
23& ${\red Qd_R}S_{3,2,1/6}S_{3,1,-1/3}$ &
24& ${\red Qd_R}S_{3,2,1/6}S_{3,3,-1/3}$ \\
25& ${\red Q}F_{3,3,-1/3}S_{3,2,1/6}S_{3,1,-1/3}$ &
26& ${\red Q}F_{3,3,-1/3}S_{3,2,1/6}S_{3,3,-1/3}$ &
27& ${\red Q}F_{3,3,-1/3}S_{3,4,1/6}S_{3,3,-1/3}$  \\
28& ${\red u_R}F_{3,2,7/6}F_{3,2,1/6}S_{3,1,2/3}$ &
29& ${\red u_R}F_{3,2,7/6}F_{3,2,1/6}S_{3,3,2/3}$ &
30& ${\red u_R}F_{3,2,7/6}F_{3,3,5/3}S_{3,2,7/6}$ \\
31& ${\red u_R}F_{3,2,7/6}S_{3,2,7/6}S_{3,1,2/3}$ &
32& ${\red u_R}F_{3,2,7/6}S_{3,2,7/6}S_{3,3,2/3}$ &
33& ${\red d_R}F_{3,2,1/6}F_{3,2,-5/6}S_{3,1,-1/3}$ \\
34& ${\red d_R}F_{3,2,1/6}F_{3,2,-5/6}S_{3,3,-1/3}$ &
35& ${\red d_R}F_{3,2,1/6}F_{3,3,2/3}S_{3,2,1/6}$ &
36& ${\red d_R}F_{3,2,1/6}S_{3,2,1/6}S_{3,1,-1/3}$ \\
37& ${\red d_R}F_{3,2,1/6}S_{3,2,1/6}S_{3,3,-1/3}$ &
38& ${\red H}F_{1,3,1}S_{1,4,3/2}$ & 
 & \\
    \hline
\end{tabular}
    }
\end{center}
    \caption{1-loop neutrino mass models for which some internal field
      can be a SM fermion. For discussion, see text.}
    \label{tab:SMLps}
\end{table}

A number of the models listed in table~\ref{tab:SMLps} have appeared
in the literature before. For example, model $\#$5 is the famous
Zee-model \cite{Zee:1980ai}. Models $\#$23 and $\#24$ are leptoquark
models \cite{AristizabalSierra:2007nf}. These are based on the idea to
break lepton number in LQ models via LQ-Higgs interactions
\cite{Hirsch:1996qy}. Note that supersymmetry with R-parity violation
generates the same diagrams \cite{Hall:1983id} with scalars that have
the same quantum numbers as in the Zee model and the LQ model $\#$23
of table~\ref{tab:SMLps}.  The particle content of models
$\#$11,$\#$12 and $\#$17, appeared first in Tables (6) and (7) of
\cite{Bonnet:2012kh}, 1-loop neutrino masses in this setup were 
discussed then in \cite{Helo:2015fba}.

Model $\#$38 is special, first because it is the only model in this 
class based on T-3. Also, while this model is technically a
``genuine'' 1-loop model in the sense, that there is no tree-level
$d=5$ neutrino mass, this model generates actually a tree-level $d=7$
mass. The model was first discussed in \cite{Babu:2009aq}. Whether
the tree-level $d=7$ or the 1-loop $d=5$ contribution is numerically
more important, depends essentially on the mass scale of $F_{1,3,1}$
and $S_{1,4,3/2}$.  If these particles are heavier than, roughly
$\Lambda \simeq 2$ TeV, the loop tends to dominate, while for lighter
masses the tree-level contribution is more important.

All models with only BSM particles in the loop are given in the 
appendix. We note, that all models in the diagram class T-I-1 
will also have a contribution to the neutrino mass matrix via 
diagram T-3. In principle, one can find all T-3 models from the 
models in T-I-1, eliminating simply the ``middle'' scalar, compare 
with figure~\ref{fig:GenDiag}. Since the number of degrees of freedom 
in T-3 and T-I-1 models are different, however, we count these 
models as different. 

Even though the size (and number) of the representations is limited 
in the exit class, very exotic states appear in our lists. For example, 
a model with $\Theta_3$ allows $SU(2)$ representations up to {\bf 6}-plets.
As one can see from the tables, there are many models that have more than 
one exit particle in the diagram. In fact, there are several models 
in which all particles in the loop are one of the particle in the 
exit lists and this is possible within any of the four diagrams. 
Since none of the fields in these tables are singlets, one can expect 
interesting phenomenology at the LHC, if the mass scale of the BSM 
particles is around the electro-weak scale. A complete study of 
possible LHC signals is, however, beyond the scope of our present 
work.

\subsection{Dark matter models\label{subsect:dm}}

\begin{figure}[t]
\centering
\includegraphics[width=0.49\textwidth]{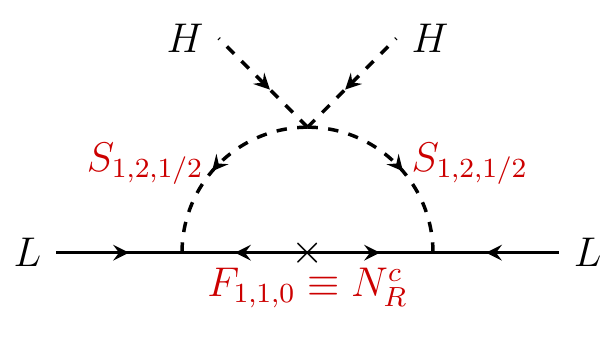}
\includegraphics[width=0.49\textwidth]{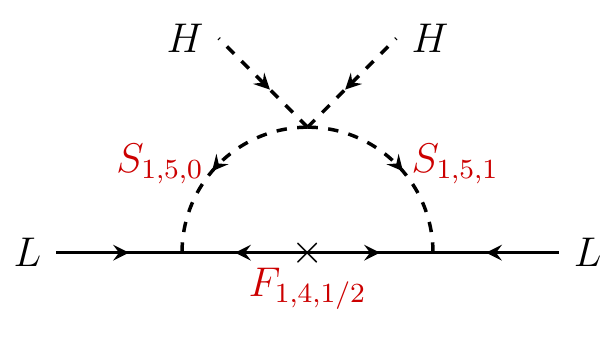}
\caption{Two examples of dark matter models. To the left the original 
scotogenic model \cite{Ma:2006km}; to the right an accidentally stable 
DM model, see text.}
\label{fig:T3DM}
\end{figure}

In this subsection we will discuss 1-loop neutrino mass models
containing a WIMP dark matter candidate. The classical proto-type for
this class of models is the scotogenic model \cite{Ma:2006km}, see
figure~\ref{fig:T3DM} (left). Models in the DM class can again be
sub-divided into two sub-classes: (i) models that need a stabilizing
(discrete) symmetry and (ii) models in which the DM candidate maybe
accidentally stable \cite{Cirelli:2005uq}, for an example see
figure~\ref{fig:T3DM} (right). We will call these two classes (i) DM-E
(since at least one of the particles in these models has quantum
numbers coinciding with one of the exit particles) and (ii) DM-A (for
accidental).

The division into these two classes can be easily understood. Consider
the scotogenic model. The particle content of this model is such, that
without any additional symmetry, beyond the gauge symmetries of the SM
a tree-level type-I seesaw would exist and the loop particles would be
unstable, i.e. they will decay to SM fields (a right-handed neutrino
can decay to SM Higgses or gauge bosons plus SM leptons, for example.).
Ref.~\cite{Ma:2006km} solves these ``problems'' with the simple
assumption that all particles in the loop transform odd under a new
$Z_2$ symmetry. The lightest of the $N_{R_i}$ and $S_{1,2,1/2}$ is
then absolutely stable. On the other hand, for larger $SU(2)_L$
multiplets as DM candidates, the model might have an accidental $Z_2$
symmetry, like in the example model shown in figure~\ref{fig:T3DM}
(right). For this model it is easy to see that $SU(2)_L$ dictates
that, at the renormalisable level, the particles in the loop always
couple in pairs to SM fields. Thus, the model has an accidental $Z_2$
and the LLP is stable automatically. We could call this an
``accidental dark matter candidate'' \cite{DiLuzio:2015oha}, in
\cite{Cirelli:2005uq} this was named ``minimal dark matter''.

Note, that this reasoning assumes that there are no other BSM fields
present in the model beyond those appearing in the 1-loop diagram.
Consider again the example model figure~\ref{fig:T3DM} (right). If we
add to this model a $S_{1,3,0}$, for example, then the vector-like
fermion can decay to a SM $L$ plus a $S_{1,3,0}$, while the latter
decays to two Higgses and this extended model will have no DM
candidate -- unless we postulate an additional symmetry, which would
put this model back into the first subclass. Thus, for the DM
candidate to be accidental DM, there should be no particle from the
exit class in the diagram and we always have to assume in our model
constructions implicitly that there are no other BSM particles in the
model, beyond those appearing in the 1-loop diagram. Otherwise, the
model will belong to DM class (i) DM-E.

However, at this point we would like to stress that for us DM-E and
DM-A are just a convenient classification scheme, dividing the model
lists into those models with ``small'' and ``large'' $SU(2)$
representations. For us, a model is specified from the particle
content and the interactions of a given 1-loop neutrino mass
diagrams. Nevertheless, any of the models may allow interactions beyond
those appearing in the diagram and those additional interactions may
put in danger the stability of the DM candidate. This issue has been
discussed in several references \cite{Cai:2016jrl,Ahriche:2016rgf,AristizabalSierra:2016vog}.
See in particular \cite{AristizabalSierra:2016vog}, where it was shown that Yukawa
interactions connecting the fermions of the neutrino mass diagram with
their vector-like partners to the scalar DM candidates, $S_{1,n,0}$,
are always possible. These interactions (together with those in the
neutrino mass diagram) induce loop decays of the DM candidate. In
essence, the stability of the DM can be interpreted as an upper limit
on this new coupling. Adding a symmetry by hand, allows to eliminate
this problem, of course, but the stability of the DM could no longer
be considered accidental.

The above arguments about the stability of accidental dark matter are
valid strictly speaking for interactions at the renormalizable
level. One can always add some discrete symmetry to a
model in our DM-A class and then our two subclasses are simply
separating models containing potential exits from models which
don't. However, in order to construct our lists of DM-A models, we
need to identify all possible WIMP DM candidates first and here
different DM-A candidates have to fulfill different theoretical
constraints. The question of which electro-weak multiplets can be good
dark matter candidates has been recently discussed in detail in two
papers \cite{Bottaro:2021snn,Bottaro:2022one}. Reference
\cite{Bottaro:2021snn} treats all possible DM candidates with $Y=0$,
while \cite{Bottaro:2022one} discusses the case $Y\ne 0$. The
following discussion draws heavily from the results of these two
papers.

Let us start with the case $Y=0$. For multiplets with $Y=0$ radiative
corrections will generate a small mass splitting among the members of
the multiplet and the sign of this splitting is such that the neutral
member of the multiplet is the lightest particle
\cite{Cirelli:2005uq}. Also, for multiplets with $Y= 0$, the neutral
member of the multiplet does not couple to the Z-boson. Thus, all DM
candidates with $Y=0$ are usually ``safe'' from existing direct detection (DD)
constraints, the best limits are currently from XENON-1t
\cite{XENON:2018voc}. However, as stressed in \cite{Bottaro:2021snn},
the situation will change in the next years and all multiplets
$(S/F)_{1,n,0}$ with $n \ge 3$ might be ruled out by future DD
experiments, such as DARWIN \cite{DARWIN:2016hyl}.

The authors of \cite{Bottaro:2021snn} then calculated the relic
density, including the effects of Sommerfeld enhancement and bound
state formation. The conclusion from this work is that all
$(S/F)_{1,n,0}$, with $n=3,\cdots,13$ can give the correct relic
density and are consistent with DD constraints. To this list, we also
have to add the singlets $(S/F)_{1,1,0}$ for completeness.

Whether these DM candidates are accidental DM or need a stabilizing
symmetry, however, depends on assumptions about possible
non-renormalizable operators and also on whether the DM is scalar or
fermion. For scalars, there always exists a $d=5$ operator of the form
${\cal O}_5^{S}\propto \frac{1}{\Lambda}S^3 H^{\dagger} H$. This
operator will lead to 1-loop DM decays and, for a coefficient of order
one, $\Lambda$ needs to be larger than the Planck scale, to guarantee
a sufficiently long-lived DM. Thus, ref. \cite{Bottaro:2021snn}
concludes that all $S_{1,n,0}$ ($n$ odd) need a stabilizing symmetry.
\footnote{Of course, unless the UV model that generates this operator
  is completely specified, the coefficient, $c$, of the operator is
  arbitrary. For $c\ll 1$ the stability of the DM looks accidental.}
For fermions, the corresponding operators are $d=6$ for $n\le 5$ or
$d=7$ for $n>5$. In particular for $n>5$ the resulting decay widths
are sufficiently small, that the DM candidate is effectively stable,
i.e. such fermions are ``good'' accidental DM, even if one takes 
into account NROs.

The discussion of the DM candidates with $Y\ne 0$
\cite{Bottaro:2022one} follows along similar lines, but is slightly
more involved. First of all, different from the case $Y=0$, for
multiplets with $Y\ne 0$ electro-weak radiative corrections do not
guarantee that the neutral member of the multiplet is the lightest
particle. In addition, multiplets with $Y \ne 0$ are strongly
constrained by DD limits, since the coupling of the DM to the Z-boson
has typically gauge strength. Thus, $Y\ne 0$ DM candidates 
must be inelastic dark matter, see below.

Again, we divide the discussion of the $Y\ne 0$ DM candidates into
different sub-cases. We start with scalars with $Y=1/2$,
i.e. $S_{1,2n,1/2}$ ($n=1,2...$).  For scalars with quantum numbers
$S_{1,2n,1/2}$ one can always write down the following quartic terms:
\begin{equation}\label{eq:pot12}
V \propto \lambda_5 (S_{1,2n,1/2}^{\dagger} 
(T^a) S_{1,2n,1/2}^c) ((H^c)^{\dagger}(\sigma^a/2) H)
+ \lambda_+  (S_{1,2n,1/2}^{\dagger} 
(T^a) S_{1,2n,1/2}) (H^{\dagger}(\sigma^a/2) H).
\end{equation}
After EWSB, these terms generate a mass splitting between the neutral
and charged components of the multiplets and for sufficiently small DM
mass one can always guarantee that the neutral state is the lightest,
by the correct choice of $\lambda_+$. Importantly, the term
proportional to $\lambda_5$ splits the neutral component of the
multiplet into its real and imaginary part. Only the lighter of the
two is the DM candidate and for sufficiently large mass splitting,
$m_0$, the DD scattering through the $Z^0$-boson diagram is
kinematically forbidden, since the coupling is of the form
$Z^0-Re(S^0)-Im(S^0)$. Such a near-degenerate system of dark matter
accompanied by a slightly heavier partner is known in the literature
as ``inelastic dark matter''. \footnote{Numerically the mass splitting
  has to be only order ${\cal O}(100)$ keV, because WIMPs move slowly
  through the galaxy.}

The scalar potential, however, also admits terms of the form
$(S_{1,2n,1/2})^2 S_{1,2n,1/2}^{\dagger} H^{\dagger}$. These terms
generate a decay width for $S_{1,2n,1/2}$ to SM Higgses and gauge
bosons at 1-loop level. Since these interactions are not suppressed by
a large energy scale, $S_{1,2n,1/2}$ can never be accidentally stable,
although it is stable at tree-level for $n>2$.

Consider next $F_{1,2n,Y}$ ($n=1,2...$). Different from the scalar
case, for fermions the mass splittings in the charged and the neutral
sector, splitting in the neutral sector, as well as decay widths,
appear only once terms at the non-renormalizable level are taken into
account, see again \cite{Bottaro:2022one}. The condition that a
sufficiently large $m_0$ be generated, to have inelastic DM, leads to
an upper bound on the new physics scale of the corresponding
operator. If this upper bound is smaller than the mass of the DM
candidate, required to give the correct relic density, the model
clearly is inconsistent. This consideration rules out all $F_{1,2n,Y}$
for $Y > 1/2$ as good DM candidates, except $F_{1,3,1}$ and
$F_{1,5,1}$.  $F_{1,2n,1/2}$, on the other hand, are good accidental
DM candidates, since the conditions of having a large enough half-live
and sufficiently large $m_0$ can be simultaneously fulfilled for 
a wide range of new physics scales. (Here, ``new'' meaning  beyond 
the fields defining the model.)

Finally, there are two more possible scalars that are consistent with
the constraints discussed above: $S_{1,3,1}$ and $S_{1,5,1}$.  Larger
representations, $S_{1,n,1}$, $n>5$, as well as larger hypercharges are
ruled out by the same argument discussed above for $F_{1,2n,Y}$.  For
$S_{1,3,1}$ and $S_{1,5,1}$ only a ``small'' window in
parameter space remains, where a large enough $m_0$ can actually be generated.
At the same time, $S_{1,3,1}$ and $S_{1,5,1}$ can decay via ``fast''
interaction terms $S_{1,3,1}HH$ and $S_{1,5,1}^{\dagger}HH(H^\dagger H)$. 
Thus, a symmetry is needed in case of $S_{1,3,1}$ and, while 
$S_{1,5,1}$ is technically in the DM-A class, making it stable 
via a symmetry is probably the preferred situation.

Technically, $F_{1,3,1}$ and $S_{1,3,1}$ belong in our list of DM-E
models, while $F_{1,5,1}$ and $S_{1,5,1}$ belongs to the class DM-A.
However, for all of these four states new physics in the ultra-violet
is needed for a consistent DM model. And the scale of this extra
physics has to exist not too far above the DM mass. This is different
from all the $Y=(1/2)$ (and $Y=0$) candidates. We thus include models
with these DM candidates in our counting, but in the appendix these
models are separated off into their own table(s).

Most importantly, multiplets larger than $n=13$ are excluded as WIMP
candidates, because the cross section necessary to reproduce the relic
density would violate unitarity bounds
\cite{Bottaro:2021snn,Bottaro:2022one}. This criterium is weaker than
the one used originally in \cite{Cirelli:2005uq}. The authors of
\cite{Cirelli:2005uq} argued that all multiplets that lead to a Landau
pole in the running of $\alpha_2$ below the GUT scale should be
excluded from the list of valid candidates.  Thus, the list of
acceptable multiplets stops at $F_{1,5,0}$ and $S_{1,7,0}$ in
\cite{Cirelli:2005uq}.  In our scanning for valid models, we follow
the weaker criterium of \cite{Bottaro:2021snn,Bottaro:2022one} and
consider multiplets up to $n=13$.

In the appendix we give the full list of models in the dark matter
class. There are four sets of tables, one for DM-E models, one for
DM-A models and another two tables for the exceptional candidates with
$Y=1$ separating again exit models and accidental DM. Note that it is impossible to separate models into lists 
for $Y=0$ and $Y=1/2$ candidates, because they mostly appear in the 
same models. Having no criterium to decide, which of the particles 
in the loop is the lightest, the models listed in the tables guarantee 
only that at least one DM candidate is present in the model, but do 
not specify which is the true DM in case there are more than one 
candidate.

Because of the $SU(2)$ contraction rule,
$n\times 2 =\{(n+1),(n-1)\}$, the models contain multiplets up to
${\bf 15}$-plets. Most of these models have Landau poles in the
evolution of the gauge couplings, at energies very close to the mass
scale of the particles in the loop, as we are going to discuss in the
next section.

\section{Renormalization group running\label{sect:rge}}

\noindent 
In the previous section we have discussed the criteria for the
construction of viable models. While there is only a finite number of
these ``phenomenologically acceptable'' models, still we found more
than 700 models in total. This number could be much further reduced,
if we apply the additional requirement, that all models remain
perturbative up to some large energy scale, say, for example, the grand
unification scale, $m_G$. Note that RGE running of gauge couplings in
a selected sample of 1-loop neutrino mass models has been studied also
in \cite{Hagedorn:2016dze}.  Perturbativity, however, is clearly a
condition expressing only a theoretical preference and thus there is a
certain amount of arbitrariness in the formulation of the exact
boundary conditions to be used. Most importantly, in the discussion in
this section we will assume that the new physics scale, where our BSM
particles live, is of the order of the electro-weak scale. Unless
noted otherwise, $m_{NP}=1$ TeV. Our motivation for this choice is
simply that for larger values of $m_{NP}$, there will not be any
chance to test the models via observables outside the neutrino sector
itself, in particular there won't be any signals at the LHC. Also, our
calculations are done at 1-loop order and thus numbers quoted below 
should be understood as rough estimates only. 

Consider the well known 1-loop solution of the renormalisation group
equations, running from a scale $t_{0}$ to $t$ is described by:
\begin{equation}
    \alpha_{i}^{-1}(t) = \alpha_{i}^{-1}(t_{0})
      -\frac{b_{i}}{2\pi} \textup{log} \left (\frac{t}{t_{0}}\right ) \, ,
\end{equation}
where $i = 1, 2, 3$ represents the $U(1)_{Y}$, $SU(2)$ and $SU(3)$
couplings and the $\beta$-functions, $b_{i}$, contain the contributions
of the SM particles plus the new fields integrated-in at the new
physics scale, denoted here as $m_{NP}=1$ TeV. Any $\alpha_{i}$ may
diverge at an energy scale $\Lambda_{LP_{i}}= t_{0} \:\textup {Exp}
\left[ \frac{2\pi}{b_{i}} \alpha_{i}^{-1}(t_{0})\right]$, which we
call the ``Landau pole''.  Clearly, with increasing $b_{i}$,
$\Lambda_{LP_{i}}$ will become lower, and larger multiplets give
larger contributions to the $b_{i}$.

\begin{figure}[t]
\centering
\includegraphics[width=0.9\textwidth]{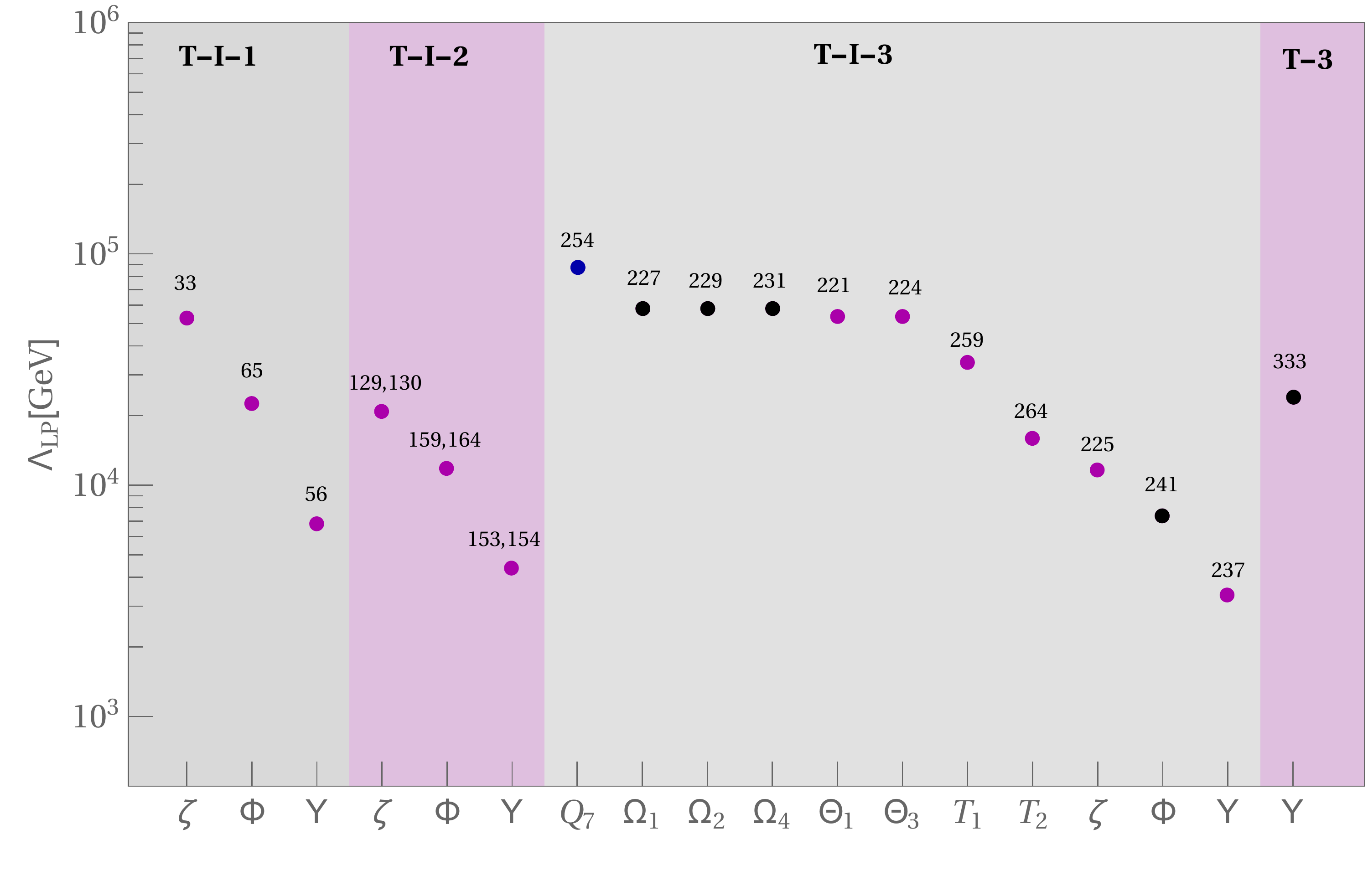}
\caption{The fields shown on the x-axis correspond to the scalar/fermionic
  exit associated with the set of models where some Landau pole 
  occurs at a low energy, $\Lambda_{LP_{i}}< 100$ TeV. The points
  represent the model(s) with the minimal LP and colours blue, magenta
  and black if such minimal LP correspond to $\Lambda_{LP_{1}}$,
  $\Lambda_{LP_{2}}$ or $\Lambda_{LP_{3}}$, respectively. The number above each
  point refers to the specific model as listed in the
  tables in the appendix. }
\label{fig:LPExits}
\end{figure}

We have calculated this running for all our models.  Let us discuss
exit models first. In tables \ref{tab:TableT11} - \ref{tab:TableT13}
in the appendix we highlighted with {\color{red}*} all models where at least one of
the $\Lambda_{LP_i}$ is small, i.e. $\Lambda_{LP_{i}} < 100$ TeV. 53
models in the exit class fullfill this criterion. Note, in this set of
low LPs models there are usually $SU(3)$-sextets, octets or
triplet(s), some of them with large hypercharges.  In figure
\ref{fig:LPExits}, we show the numerical value for the Landau poles
for a number of models. Here, we show only those models which have the
lowest $\Lambda_{LP_{i}}$ for a given exit particle.  The numbers
above the points refer to the specific model following the enumeration
in the tables in the appendix. The topology containing most of the
models with low LPs is T-I-3 (since it contains three fermions) while
T-3 has the fewest ``endangered'' models. Of all the models, there is
only one, where the minimal LP occurs in $\alpha_1$
($\Lambda_{LP_1}$).

In the tables in the appendix we have also marked all models, in which
no Landau pole appears below the GUT scale (chosen to be $m_G \simeq
10^{15}$ GeV). There are a total of 57 of such ``safe'' models in the
exit class. Thus, applying this strict criterion, would allow 
eliminating 87 \% of our 368 exit models.

\begin{figure}[t]
\centering
\includegraphics[width=0.7\textwidth]{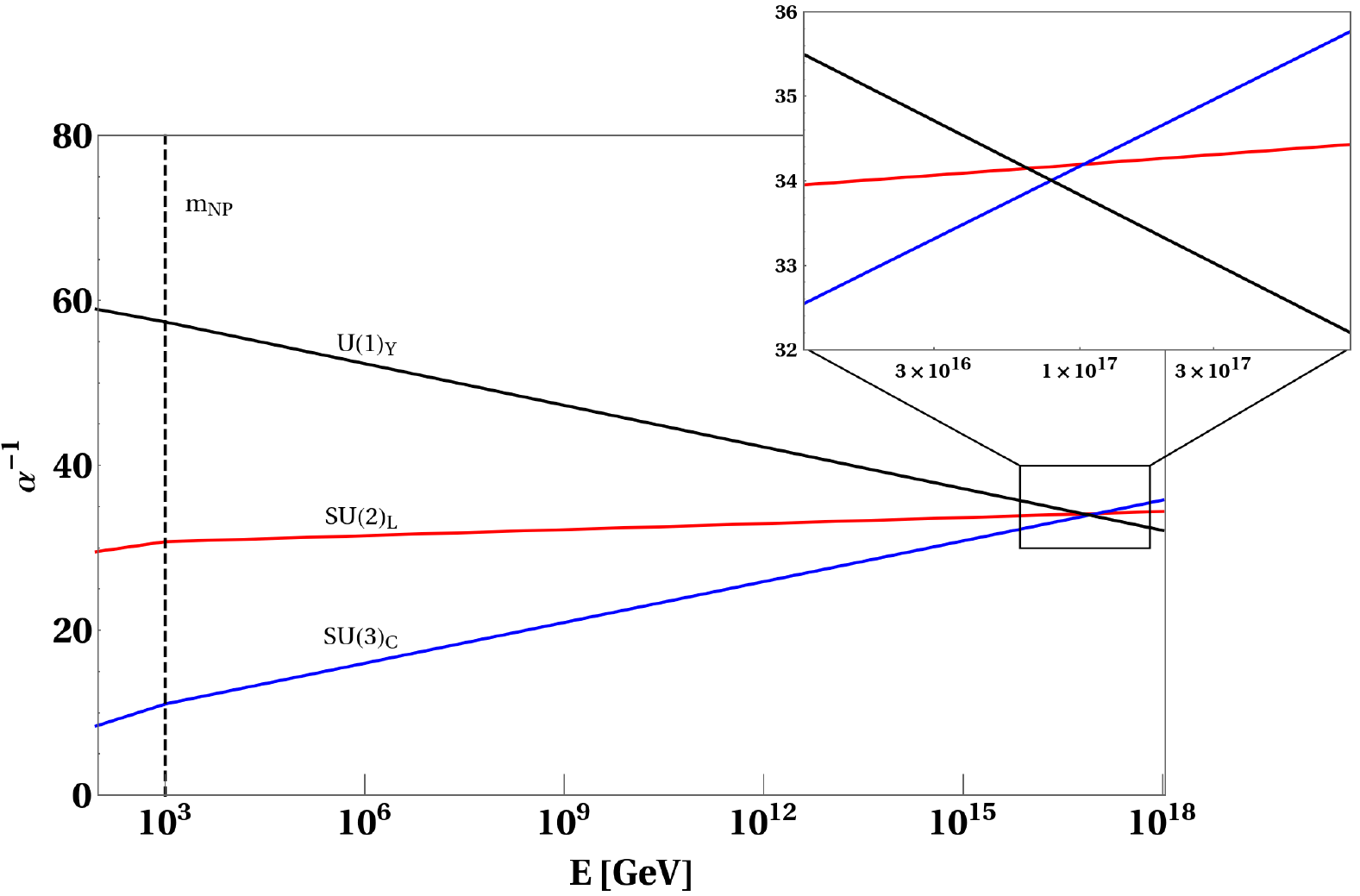}
\caption{Running of the gauge couplings for the model $(D,Q_{1},
  \Pi_{1}, \omega_{1})$ from the T-I-2 class. This model unifies at a
  scale of $m_{G} \simeq 10^{17}$. In this plot all BSM fields are added at
  a scale of $m_{NP}=1$ TeV.}
\label{fig:GCUTI2}
\end{figure}

We would also like to stress, of all the 368 models containing scalar
or/and fermionic exit, there is only one, where the gauge couplings
unify at a scale large enough to avoid constraints from proton
decay. Figure~\ref{fig:GCUTI2} shows the running of the gauge
couplings for this model. The model is marked in
table~\ref{tab:TableT12} with a purple dagger
{\color{violet}$\boldsymbol{\dagger}$}, it is model number 211. It
belongs to diagram T-I-2 and contains 2 scalar and 2 fermionic exits:
$(D, Q_{1},\Pi_{1}, \omega_{1})$. We note in passing that we have
found 8 additional models, in which we observe a quantitatively
acceptable unification of the gauge couplings, but at a rather low
scale, where one would expect to have too short proton decay
half-lives. Note however, that the authors of \cite{Evans:2022dgq}
recently discussed a mechanism to sufficiently suppress proton decay
rates, even if the unification scale is below our (conservative)
cutoff of $m_G = 10^{15}$ GeV. We have not worked out, whether such a
suppression would work quantitatively in any of our 8 low-unification
models, since it depends on the details of the GUT model building,
which is beyond the scope of our present work.

For completeness, we also checked the 38 models containing SM
fermions, see table~\ref{tab:SMLps}. Of these, 29 models do not have
Landau poles below $m_G = 10^{15}$ GeV. Also, in this class there are
4 models that give quantitatively good unification of gauge couplings
but in all cases at energies below $10^{15}$ GeV.

\begin{figure}[t]
\centering
\includegraphics[width=0.49\textwidth]{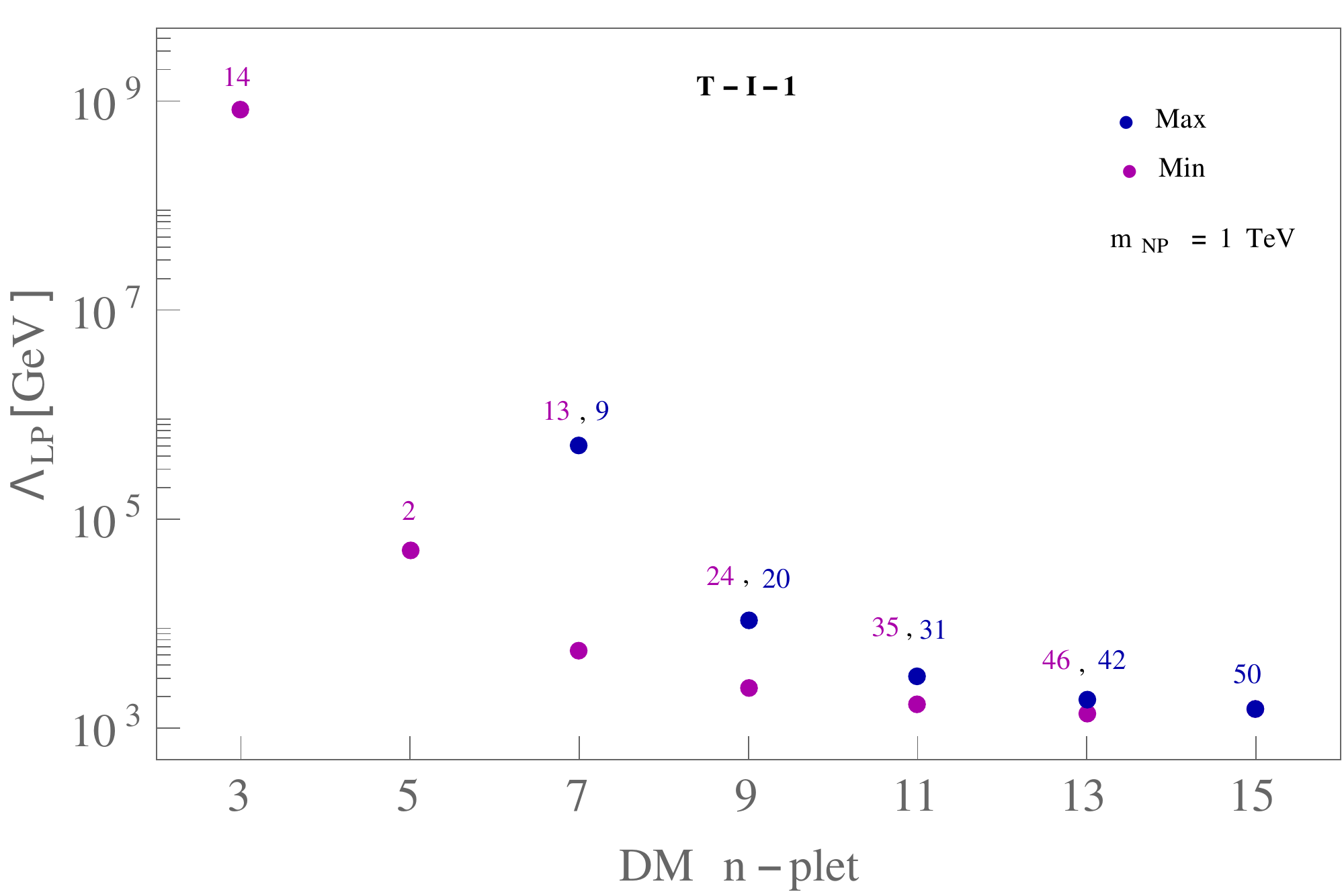}
\includegraphics[width=0.49\textwidth]{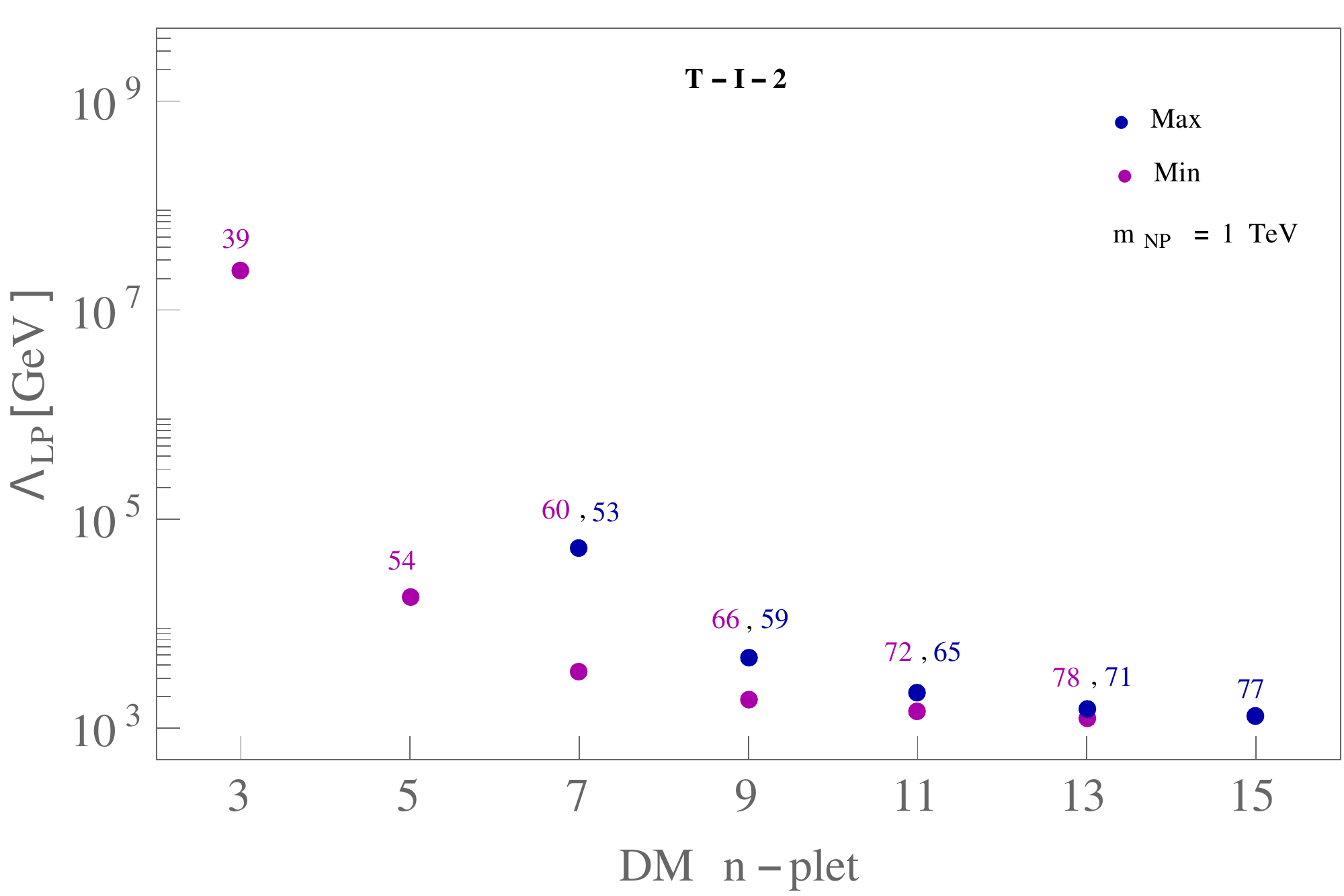}
\\
\includegraphics[width=0.49\textwidth]{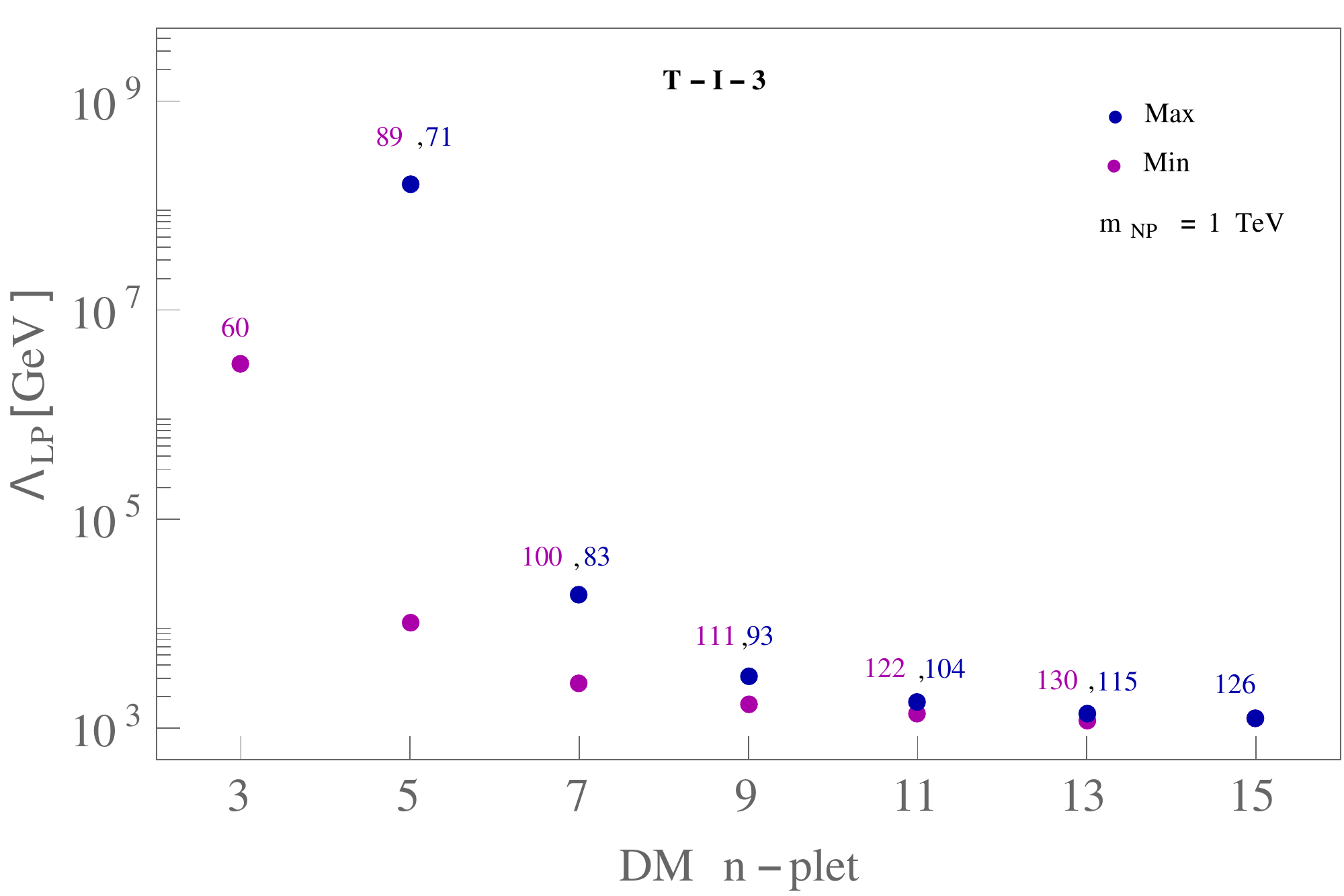}
\includegraphics[width=0.49\textwidth]{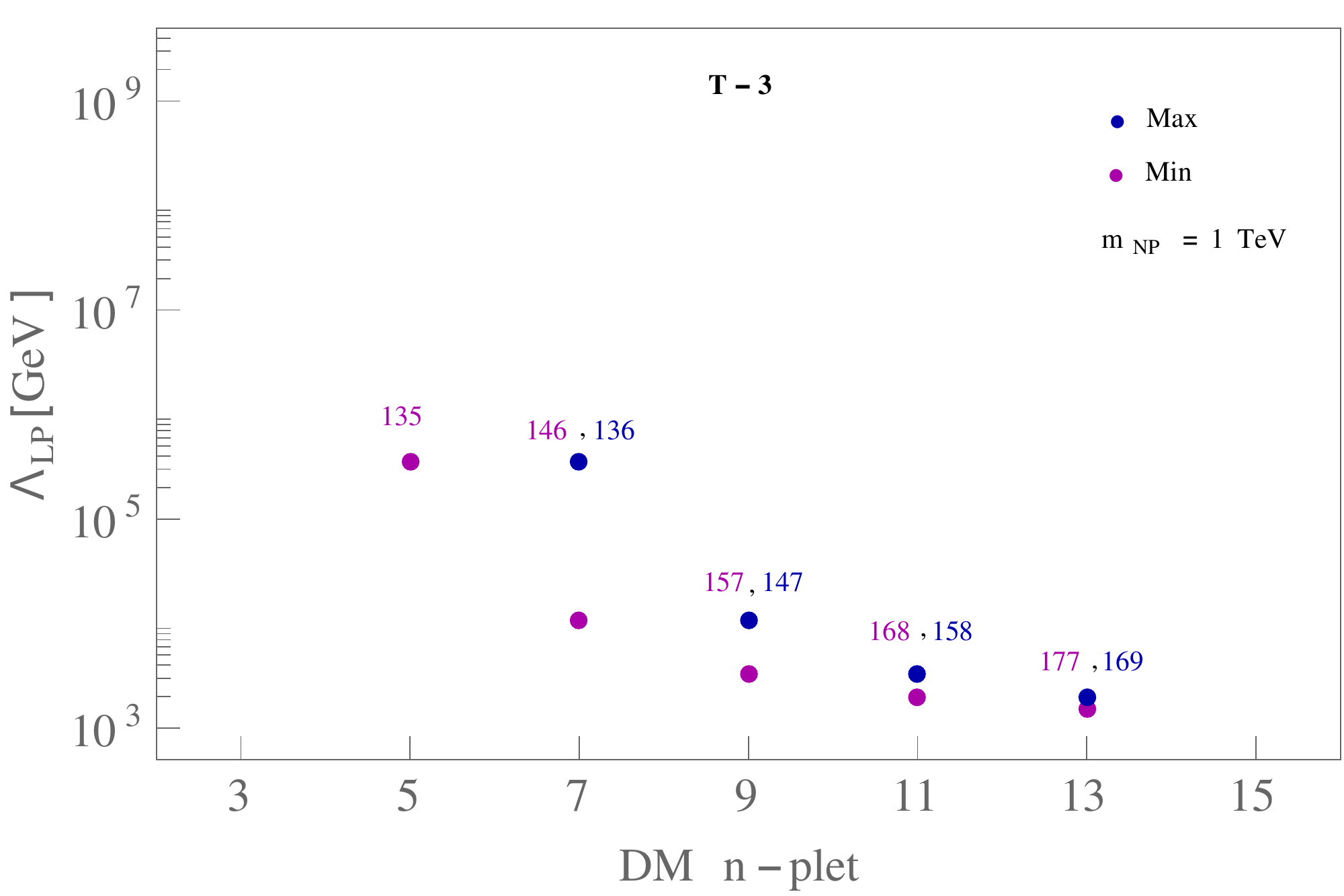}
\caption{For each topology and electro-weak DM representation, we show
  the models with the minimal (magenta points) and maximal (blue
  points) $SU(2)$-Landau pole $\Lambda_{LP_{2}}$ and its specific
  values. In the running of the gauge couplings all the additional
  fields beyond the SM are added at the scale $m_{NP}=1$ TeV. The
  numbers above the points refer to the specific model as listed in
  the DM tables in the appendix. All model numbers refer to DM-A in
  table \ref{tab:TableDM-A}, except the ones associated to 3-plet DM
  (and also model 71 of T-I-3) which belong to DM-E in table
  \ref{tab:TableDM-E}.}
\label{fig:LPDM}
\end{figure}

Now we turn to models containing a DM candidate, shown in tables
\ref{tab:TableDM-E}-\ref{tab:TableDM-Ex} of the appendix.  As could be
expected, models with large $SU(2)$ multiplets DM fields lead also to
Landau poles $\Lambda_{LP_{2}}$ fairly below the GUT scale, as
depicted in figure~\ref{fig:LPDM}. In this figure, we show only the
``extreme'' models, i.e. those in which the LP scale takes either its
minimal or maximal value for a given DM $n$-plet. (All other models
for that DM $n$-plet have LPs in between the values shown). Large DM
representations, i.e. $n \geq 7$, lead $\Lambda_{LP_{2}}$ values in the
ballpark of a few TeVs for almost all topologies. Choosing a larger
$m_{NP}$ would give also larger values for $\Lambda_{LP_{i}}$. Of
course, however the LPs are always only a factor of a few above the new
physics scale.

For T-3, fewer new particles are needed, resulting in smaller $b_{i}$
coefficients. Thus, the LPs appear at relatively larger energies.  If
we were to impose the criterion of having scenarios free of LPs below
$m_G$, the list of acceptable models would stop at $n=5$, thus
eliminating most of the DM models. However, as discussed in section 2,
in our scanning for valid models, we followed the weaker criterion of
\cite{Bottaro:2021snn}.

Also in the DM class we found some models
which unify the gauge couplings. In total, there are 8 DM models with
a quantitatively acceptable gauge coupling unification, but in all of
these models the unification scale lies below $m_{G}<10^{15}$ GeV.

\begin{table}
\begin{center}
  \begin{tabular}{  c | c | c }
   DM $n$-plet & $M_{\chi}$ & Min $\Lambda_{LP_{2}}$\\ \hline 
    3 & 2.5 TeV & $3 \times 10^{6}$ TeV \\ \hline
%    3 & 2.53 TeV & $2.64 \times 10^{6}$ TeV \\ \hline
    5 & 15.4 TeV & $9 \times 10^{2}$ TeV \\ \hline
%    5 & 15.4 TeV & $9.38 \times 10^{2}$ TeV \\ \hline
    7 & 54.2 TeV & $3 \times 10^{2}$ TeV \\\hline
%    7 & 54.2 TeV & $3.29 \times 10^{2}$ TeV \\\hline
    9 & 117.8 TeV & $3 \times 10^{2}$ TeV \\\hline
%    9 & 117.8 TeV & $3.06 \times 10^{2}$ TeV \\\hline
    11 & 199 TeV & $4 \times 10^{2}$ TeV \\\hline
%    11 & 199 TeV & $3.50 \times 10^{2}$ TeV \\\hline
    13 & 338 TeV & $5 \times 10^{2}$ TeV 
%    13 & 338 TeV & $4.86 \times 10^{2}$ TeV 
\end{tabular}
\caption{$SU(2)$ $n$-plet WIMP thermal masses, for which the relic
  abundance of dark matter would be correctly reproduced and energy
  scale of the Landau pole, $\Lambda_{LP_{2}}$. In the running of the
  gauge couplings, all the extra fields up to the SM are added at the
  $M_{\chi}$ scale.}
\label{tab:RDmasses}    
\end{center}
  \end{table}

For the running of the gauge couplings, we have considered all the
extra fields, including the DM candidates, to have a common mass of
$m_{NP} = 1$ TeV. For DM models this choice may not be completely
realistic in the following sense: For a given representation one can
calculate a mass scale, at which the chosen DM candidate would have
the correct relic density to reproduce the measured DM abundance
\cite{Bottaro:2021snn}.  If we chose those values of the DM mass as
$m_{NP}$, we obtain different numerical values for $\Lambda_{LP_{2}}$.
In table~\ref{tab:RDmasses} we show the corresponding values of the
WIMP thermal masses ($M_{\chi}$), taken from \cite{Bottaro:2021snn},
for each DM representation and also the LP values for topology
T-I-1. (For the rest of the topologies, the value of the LP is simply
re-scaled by a multiplicative factor with respect to the values of
figure~\ref{fig:LPDM}.) The numerical values in the table differ, 
of course, from the values shown in the figure, but the principle 
observation remains unchanged: Landau poles appear at relatively 
low energies and thus most DM models would be excluded, if we 
require perturbativity up to the GUT scale. 

In summary, the requirement that models remain perturbative up to 
some large energy scale (GUT scale) would exclude a large number 
of models from both, the exit and the DM class, if the new physics 
scale is of the order of $m_{NP} \sim 1$ TeV.

\section{Discussion\label{sect:sum}}

In the current paper we have attempted to give an answer to the
question: How many 1-loop models for neutrino masses can be
constructed at $d=5$?  We have presented complete lists of models
based on certain selection criteria. Given the assumptions spelled
out in detail in section \ref{sect:setup}, we have found a total of
724 1-loop neutrino mass models: 406 in the exit class and 318 in the
dark matter class. While these are certainly uncomfortably big
numbers, especially compared to the fact that there are only three
tree-level seesaws, many of these models could actually be excluded in
the future.

For the dark matter class, future DD experiments, such as DARWIN
\cite{DARWIN:2016hyl}, will either finally detect WIMP dark matter 
or exclude most of the larger $SU(2)$ multiplets
\cite{Bottaro:2021snn,Bottaro:2022one} as DM candidates. From our 
318 DM models only 109 would survive non-observation of DM in 
DARWIN.

Also, there are theoretical considerations, such as perturbativity up
to some large energy scales, that we have discussed in section
\ref{sect:rge}. Conservatively, we have listed all possible models in
our tables. However, if we require our new physics scale, at which the
1-loop neutrino mass is generated, to be around the electro-weak scale
and add the condition that all gauge couplings remain perturbative up
to the GUT scale, only 57 models (out of 406) in our exit class
survive. Similarly, in the DM class only 59 out of the whole 318
would survive this constraint, eliminating in particular all models
with representations larger than {\bf 5}-plets.

Two important assumptions on model building were used in all our
constructions: (i) use only scalars and fermions as BSM fields; and
(ii) avoid stable charged relics. Both of these assumptions can be
questioned. Let us discuss 1-loop models with vectors first.  Note
that very few 1-loop models with gauge vectors do exist in the literature, a
few examples are \cite{Ma:2012xj,Deppisch:2016qqd,Fonseca:2016xsy}.
The two main problems with gauge vectors are that: (a) For many of the
vectors, which appear in the 1-loop diagrams, it is not even possible
to find a phenomenologically consistent or interesting gauge group
\cite{Fonseca:2016jbm}; and (b) complete gauge models in many cases
also contain the ingredients for a tree-level seesaw, thus loops 
are most likely only a sub-dominant contribution to the neutrino 
mass in these constructions.

\begin{figure}[t]
\centering
\includegraphics[width=0.44\textwidth]{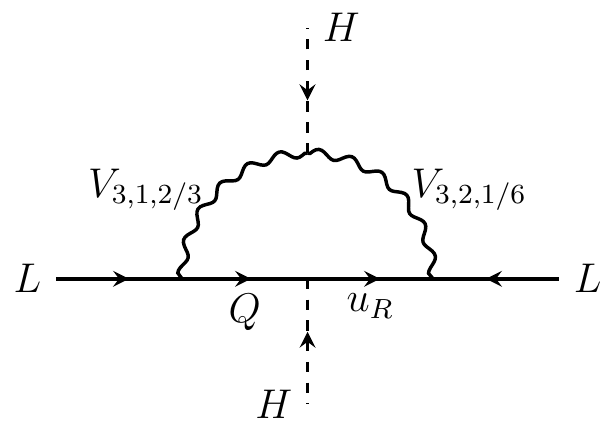}
\includegraphics[width=0.44\textwidth]{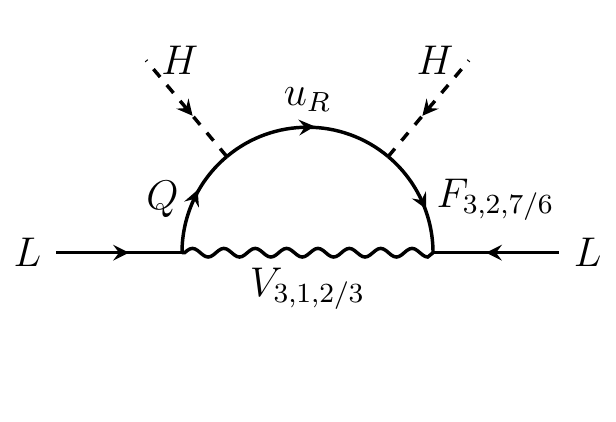}
\caption{Two example 1-loop diagrams with vectors instead of
  scalars. For a discussion see text.}
\label{fig:VecExample}
\end{figure}

Disregarding these problems in the construction of valid gauge
models, however, our automated diagram-based approach allows us,
of course, to search also for valid 1-loop neutrino mass diagrams
with vectors instead of scalars. From the list of valid
``exit'' vectors, see table (3) of \cite{deBlas:2017xtg}, one
can show that there are a total of 499 vector models in the
exit class, out of which 34 models contain either one or two
SM fermions. Two examples are shown in figure~\ref{fig:VecExample}.
The example on the left is from the diagram class T-I-2, while
the one on the right is from T-I-3. Both diagrams have vector LQs 
as internal particles. Let's have a closer look to
the diagram on the left first. The quantum numbers of the vectors
are the same as in the scalar LQ model of table~\ref{tab:SMLps},
model $\#$23, except the hypercharge $Y=2/3$ of the second vector
in the diagram. Thus, the diagram contains $u_R$ instead of $d_R$,
but is otherwise very similar to the corresponding scalar LQ model.
Many, but not all of our scalar models can be ``vectorized'' by such
simple replacements. 

The two examples shown in figure~\ref{fig:VecExample} can also serve
to discuss the main problems one encounters in the construction of
neutrino mass models based on extended gauge theories. Consider the
model shown in the figure on the right. The vector in this diagram,
$V_{3,1,2/3}$, can be generated from the adjoint of $SU(4)$, when the
Pati-Salam (PS) group \cite{Pati:1974yy}, is broken to the standard
model. Both, $Q$ and $u_R$ are of course present in Pati-Salam as
members of the ${\bf 4}$ and $\overline{\bf 4}$. However, the
$F_{3,2,7/6}$ is not part of a minimal PS model and thus another
multiplet containing this fermion has to be added to complete the
particle content of this 1-loop model. On the other hand, the
$\overline{\bf 4}$ necessarily contains a $F_{1,1,0}$, i.e. a
right-handed neutrino. Since the 1-loop diagram necessarily violates
lepton number, it seems reasonable that the model also generates a
Majorana mass term for $N_R$. This could, for example, occur if the PS
is first broken to the left-right group, which is then broken by a
right-triplet to the SM group. The model then could generate the
1-loop diagram shown, but also has a tree-level seesaw to which the
loop diagram would be only a minor correction in large parts of the
parameter space. In other words, according to our criteria, in this
setup the 1-loop diagram would not be considered ``genuine''. This
problem -- the presence of $N_R$ or also other tree-level seesaws --
occurs in many of the popular gauge groups, in which the SM group
could be embedded.  While it seems possible to construct a full model
along the lines just discussed, in which the tree-level seesaw is
sub-dominant (or absent entirely at tree-level), the model building
required clearly is beyond our minimalistic approach to neutrino
masses.

A second problem with vector diagrams is demonstrated by the diagram
on the left of figure~\ref{fig:VecExample}. Here, no BSM fermion
appears, but two different vectors are needed to complete the
diagram. Again, $V_{3,1,2/3}$ appears in the adjoint of $SU(4)$. The
other vector, $V_{3,2,1/6}$, appears, for example, in flipped $SU(5)$
\cite{Barr:1981qv}.  Flipped $SU(5)$ has no $N_R$, so no tree-level
seesaw type-I, but it induces proton decay and the ``vector
leptoquark'' $V_{3,2,1/6}$ of the 1-loop diagram has also diquark
couplings in this setup. Thus, when $V_{3,2,1/6}$ is interpreted as
the gauge vector of flipped $SU(5)$, its mass must lie at the grand
unification scale.  This mass scale is too large to generate the
atmospheric neutrino mass scale with perturbative couplings from a
1-loop diagram. The problem is exacerbated by the fact, that the
diagram needs two vectors. Thus, one would need to identify a group --
or semisimple group \cite{Allanach:2021bfe} -- which contains both,
the Pati-Salam group and the $SU(5)$, plus suitable model building to
avoid proton decay and many other constraints in this extended theory.
Finally, we would like to stress again, that for most of the exit
vectors \cite{deBlas:2017xtg} it was shown in \cite{Fonseca:2016jbm},
that no suitable gauge group can be constructed at all, since they can
not lead to models which contain the SM particle content.

Our second main assumption is to avoid stable charged relics.  For the
models in the exit class, one can actually question the validity of
this criterion. From experimental data the absence of stable charged
particles is established only for a certain mass window, roughly $M
\sim [1, 10^5]$ GeV \cite{ParticleDataGroup:2020ssz,Hemmick:1989ns}.
1-loop models for neutrino mass, however, can fit the observed data
even for considerably more massive BSM states in the loop, roughly up
to $10^{(12-13)}$ GeV for perturbative couplings. Such ultra-heavy
particles would decouple very early in the history of the universe and
therefore not be produced in any measurable quantities.
\footnote{It is even conceivable such states are not produced at all,
if the reheat temperature of the universe is sufficiently below
  the mass of these BSM states.} Thus, there is a window of parameter
space for 1-loop neutrino mass models, where this criterion is not
supported by experimental data. Clearly, we have disregarded
this possibility. We note in passing, that such models would use, 
of course, even larger multiplets than what we have considered and
thus Landau poles would exist in these constructions always not far
above the mass scale of the BSM states.

For the dark matter models, on the other hand, the two most important
constraints for valid WIMP candidates are (i) unitarity bounds on the
annihilation cross section in the early universe and (ii) limits by
direct detection experiments. Here we followed \cite{Bottaro:2021snn}
and \cite{Bottaro:2022one}. While unitarity bounds put a definitive
upper limit on the size of the $SU(2)$ multiplet, that can be a good
WIMP candidate, the argument (ii) is slightly more fragile. We have 
considered models with $Y=0$ DM candidates, as well as inelastic 
DM candidates. However, one could think about cooking up other 
ways to avoid the DD constraints and we have simply disregarded this 
possibility.

In summary, we provide ``complete'' lists of possible 1-loop models
for neutrino masses. We have considered two possible classes of models,
which can be consistent with cosmology: ``Exit'' models, with no
stable particles in the loop and dark matter models, which assume
that the lightest particle in the loop is neutral, stable and can
be in agreement with known constraints. In the appendix we give
the lists of all possible models, consistent with these assumptions.
It would be interesting to study, whether some of these models can
lead to phenomenology at colliders, say the LHC or FCC, that has not
already been covered in previous work, see for example 
\cite{Cai:2017jrq,AristizabalSierra:2007nf,Cai:2014kra,Ghosh:2017jbw,%
Nomura:2016ask,Khan:2018jge,Hirsch:2019gge,R:2020odv,Avnish:2020rhx,%
Gargalionis:2019drk,Ashanujjaman:2022cso}.

\section{Complete lists of 1-loop neutrino mass models\label{sect:appendix}}

Here we give tables containing the 1-loop neutrino mass models as discussed in the previous sections. The models are divided in tables for each of the four 1-loop neutrino mass diagrams: T-I-1, T-I-2, T-I-3 and T-3, see figure~\ref{fig:GenDiag}. For each diagram, the models are classified into two large classes: ``exit'' (tables \ref{tab:TableT11}, \ref{tab:TableT12}, \ref{tab:TableT13}) and dark matter (tables \ref{tab:TableDM-E}, \ref{tab:TableDM-A}, \ref{tab:TableDM-Ex}) models.  The exit models  have been ordered in the tables from top to bottom  by models with 1 exit to models with 4 exits. To identify the exits particles we have used the notation of Ref. \cite{deBlas:2017xtg}, shown in tables~\ref{t:scalars} and \ref{t:fermions}. The DM models have been separated in four class of models: Models with exits that need a stabilizing symmetry to give an acceptable DM candidate (DM-E: table \ref{tab:TableDM-E}), models in which the DM is stable due to an accidental symmetry (DM-A: table \ref{tab:TableDM-A}) and another two cases for exceptional candidates with $Y = 1$ which are separated again in exit DM models  (DM-E exceptional: table \ref{tab:TableDM-Ex})  and accidental DM models (DM-A exceptional: table \ref{tab:TableDM-Ex}). See section \ref{sect:setup} for discussion.

The symbol {\color{red}*} has been placed next to each model where one of the Landau pole scales is very low, i.e $\Lambda_{1,2,3} < 100$ TeV. On the other hand the symbol ``$-$'', placed next to a model, represents models where all Landau pole scales, larger than $m_{NP}$, are very large $\Lambda_{1,2,3} > 10^ {15}$ GeV. The symbol {\color{violet}$\boldsymbol{\dagger}$} is marked next  to the only model that unifies at a scale of $m_{G} \simeq 10^{17}$ GeV.

\begin{table}[H]
\includegraphics[width=1.1\textwidth]{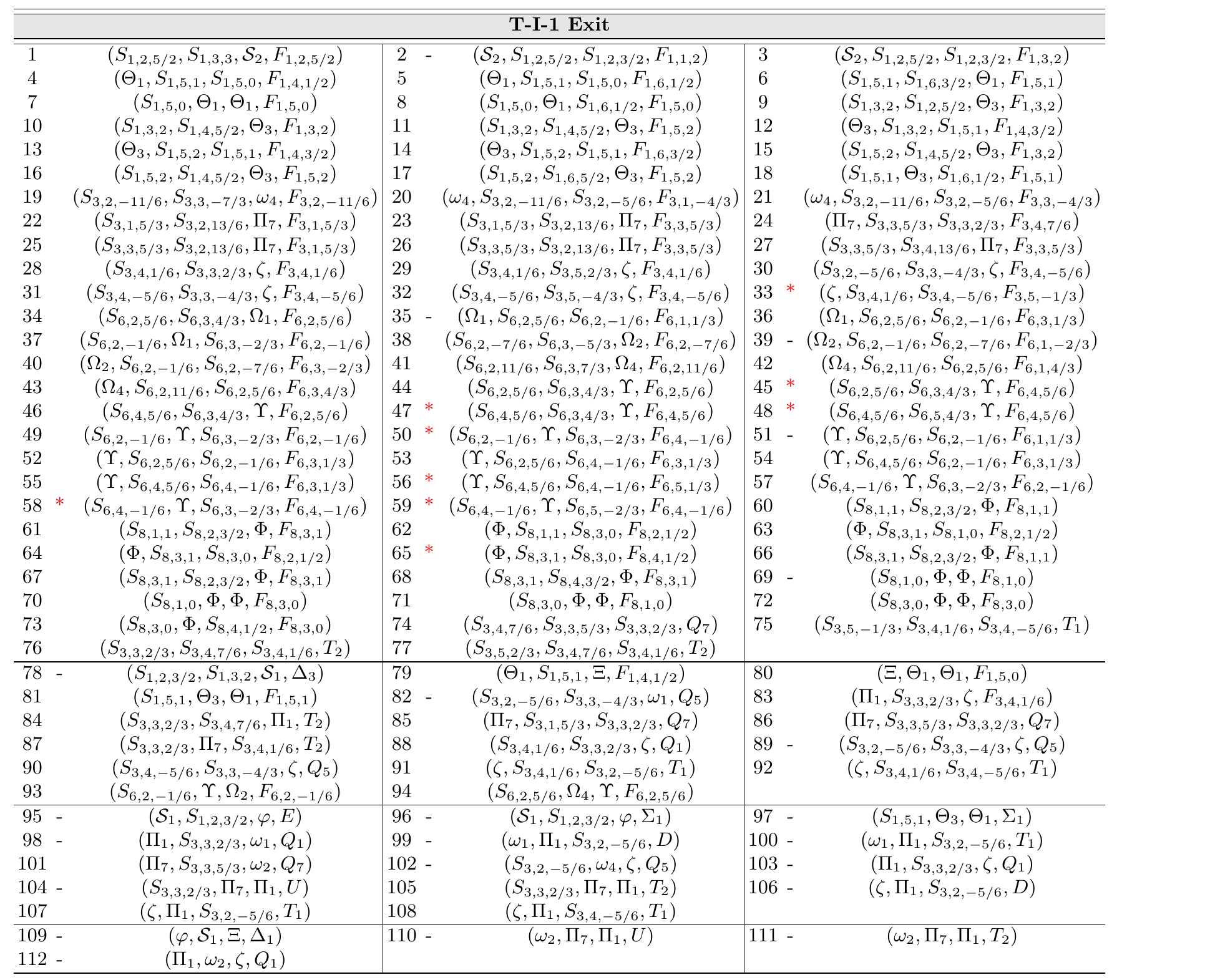}
    \caption{T-I-1 models with scalar and fermionic exits. The four horizontal divisions represent the models with 1, 2, 3 and 4 exit fields. The {\color{red}*} and ``$-$'' represent the models where one of the Landau pole scales is very low, i.e. $\Lambda_{1, 2, 3} < 100$ TeV, and where there is no Landau pole up to $10^ {15}$ GeV.}
    \label{tab:TableT11}
\end{table}

\begin{table}[H]
\includegraphics[width=1.1\textwidth]{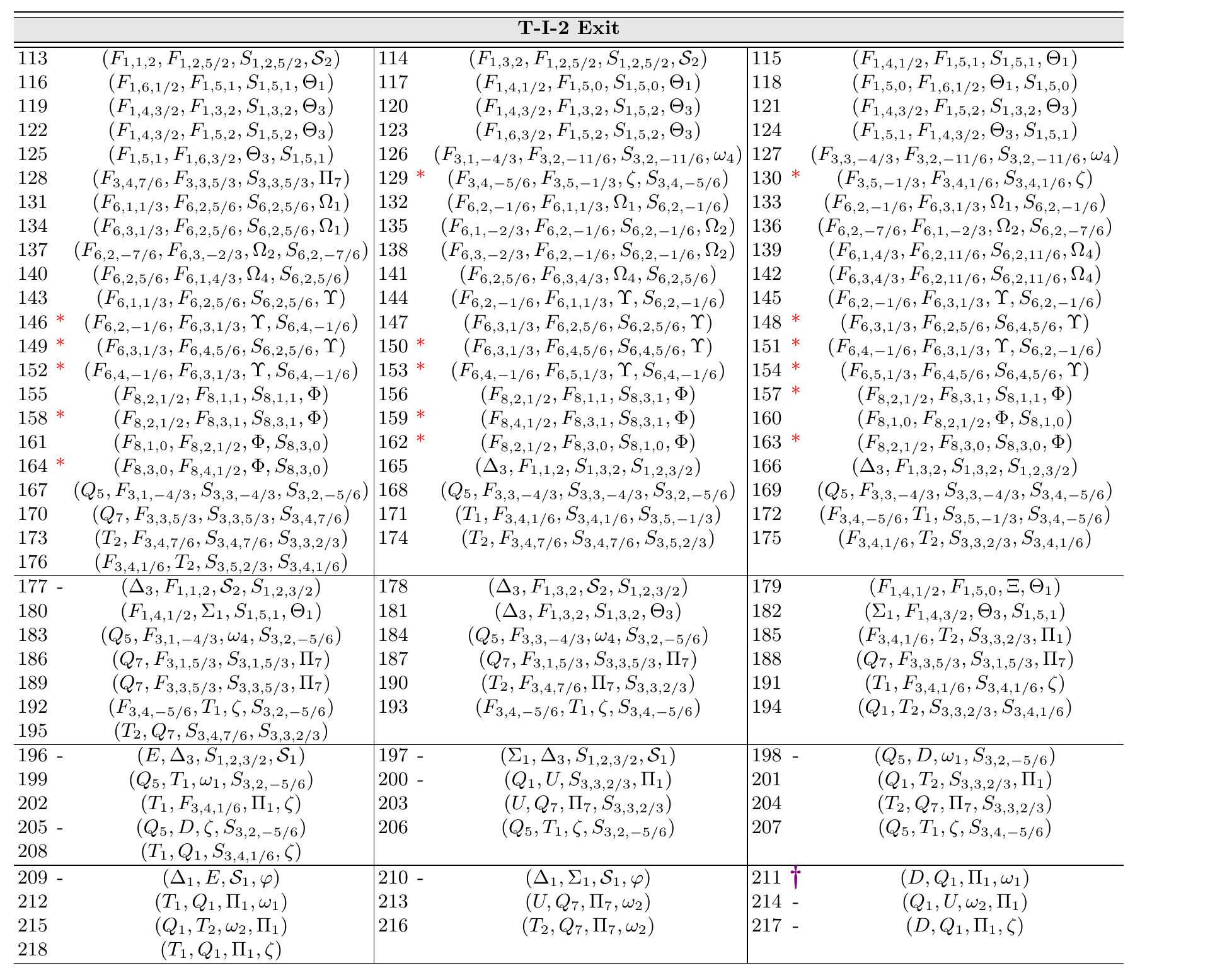}
\caption{T-I-2 models with scalar and fermionic exits: The model {\color{violet}$\boldsymbol{\dagger}$} unifies at a scale of $m_{G} \simeq 10^{17}$ GeV .}
\label{tab:TableT12}
\end{table}

\begin{table}[H]
\includegraphics[width=1.1\textwidth]{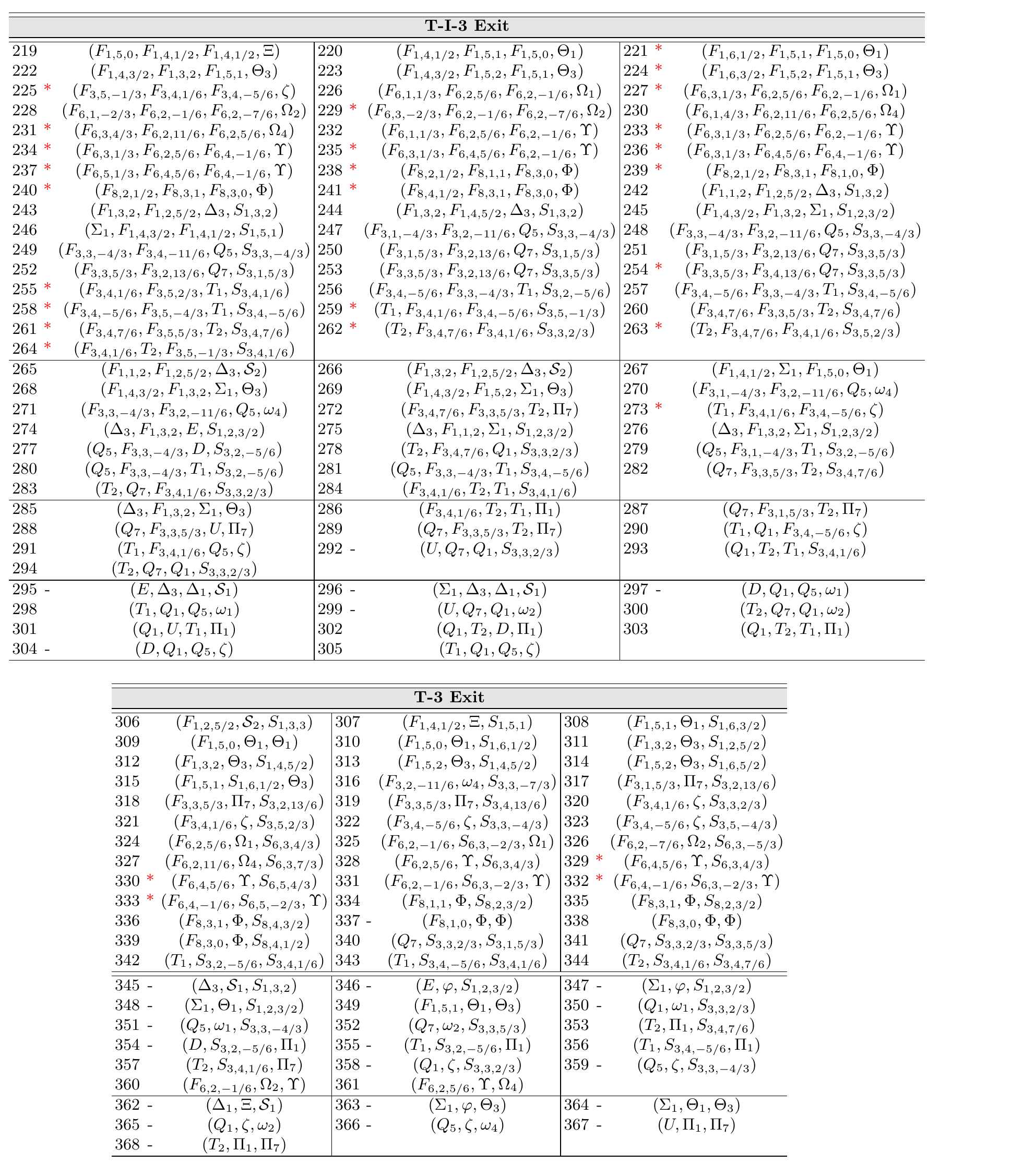}
\caption{T-I-3 and T-3 models with scalar and fermionic exits. }
\label{tab:TableT13}
\end{table}

\begin{table}[H]
\includegraphics[width=1.1\textwidth]{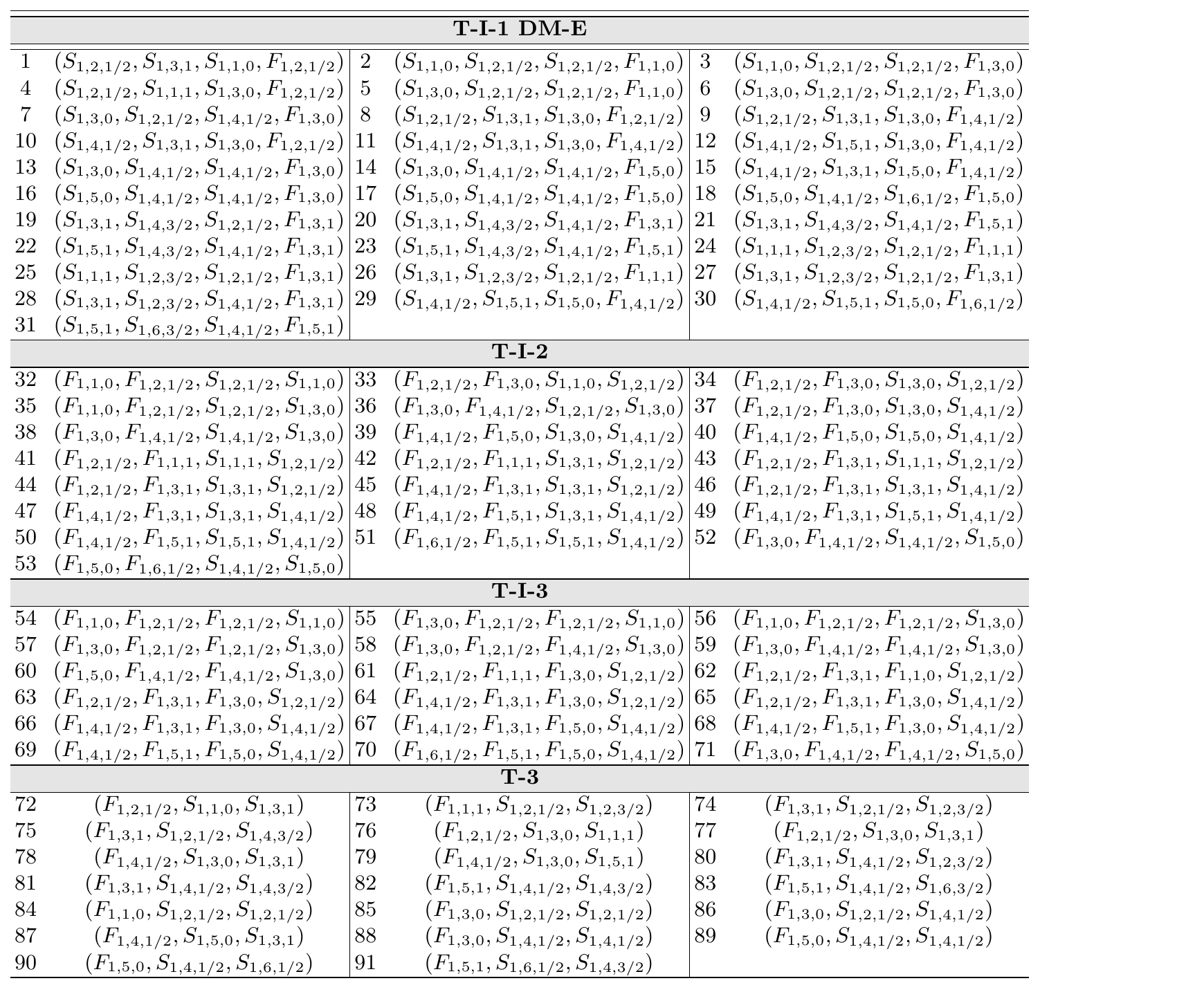}
\caption{DM models with exits (DM-E) which need a stabilizing symmetry to give an acceptable WIMP candidate.}
\label{tab:TableDM-E}
\end{table}

\begin{table}[H]
\includegraphics[width=0.92\textwidth]{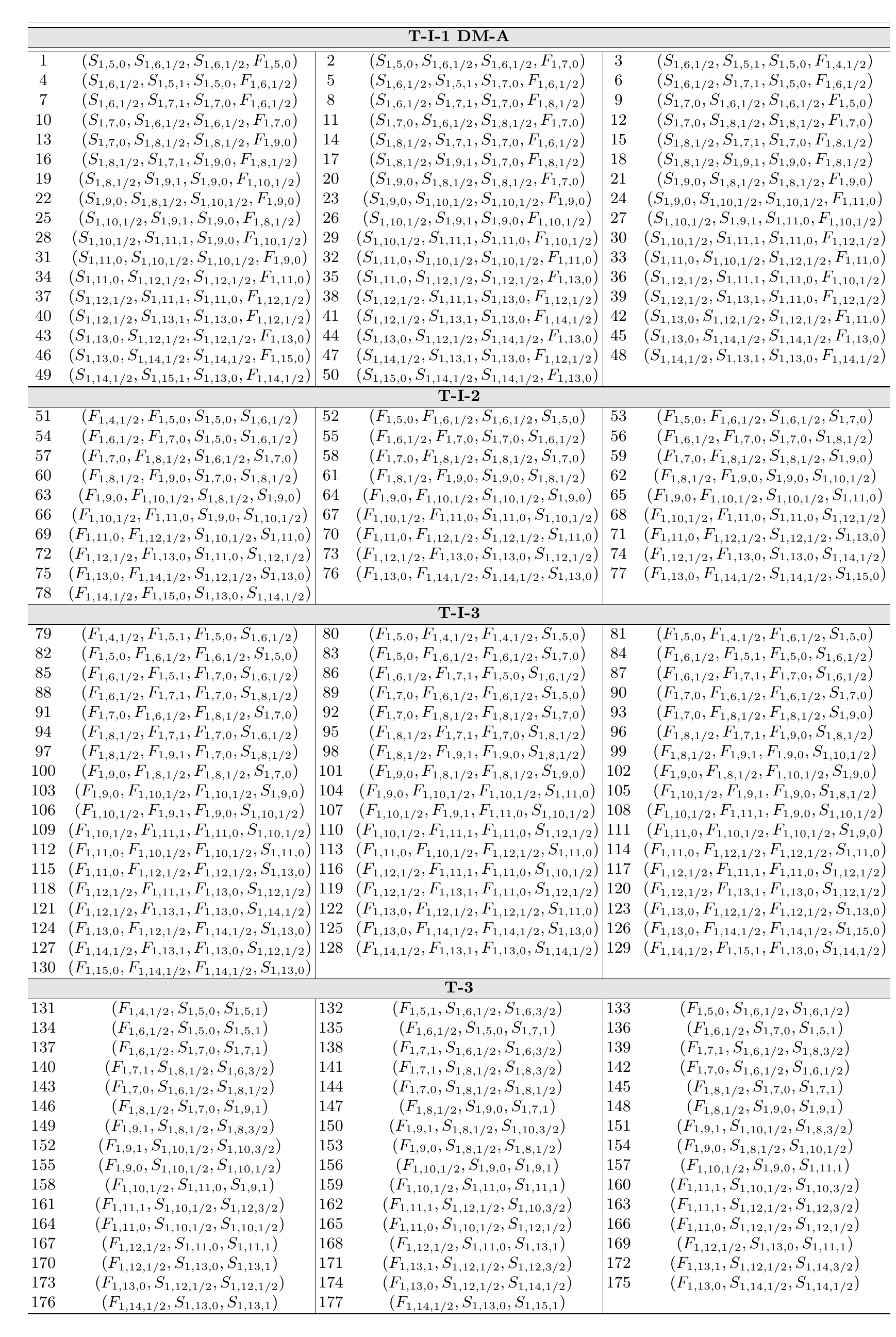}
\caption{Models in which the DM could be stable due to accidental symmetry (DM-A).}
\label{tab:TableDM-A}
\end{table}

\begin{table}[H]
\includegraphics[width=1.1\textwidth]{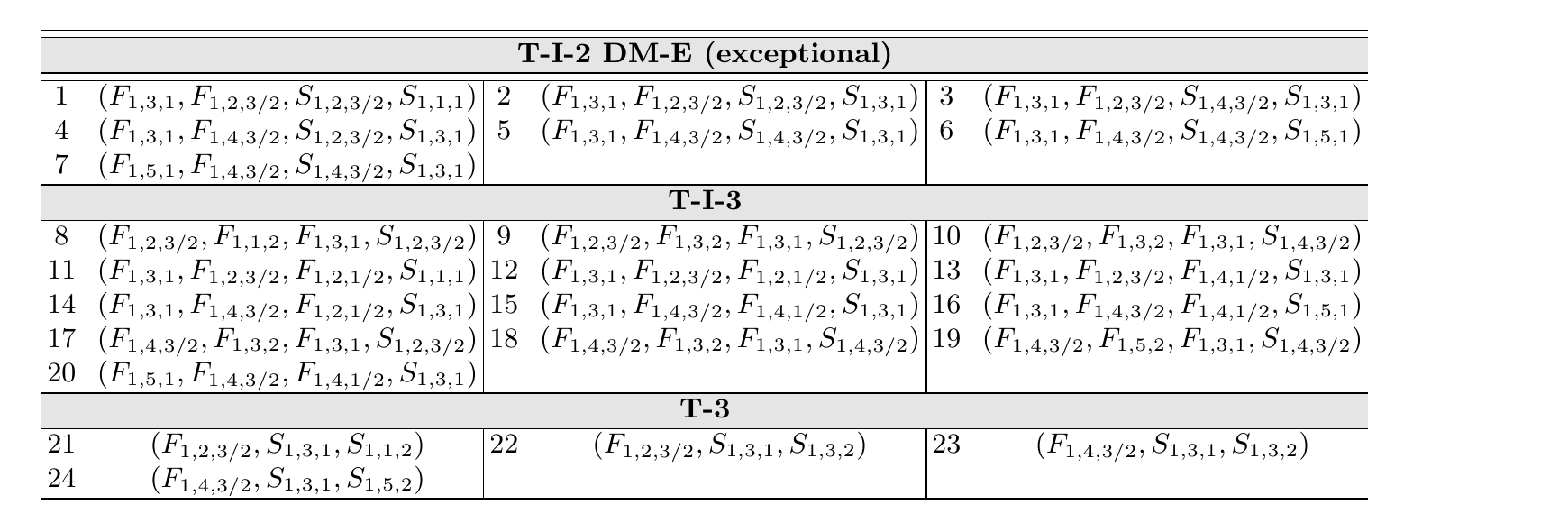}
\includegraphics[width=1.1\textwidth]{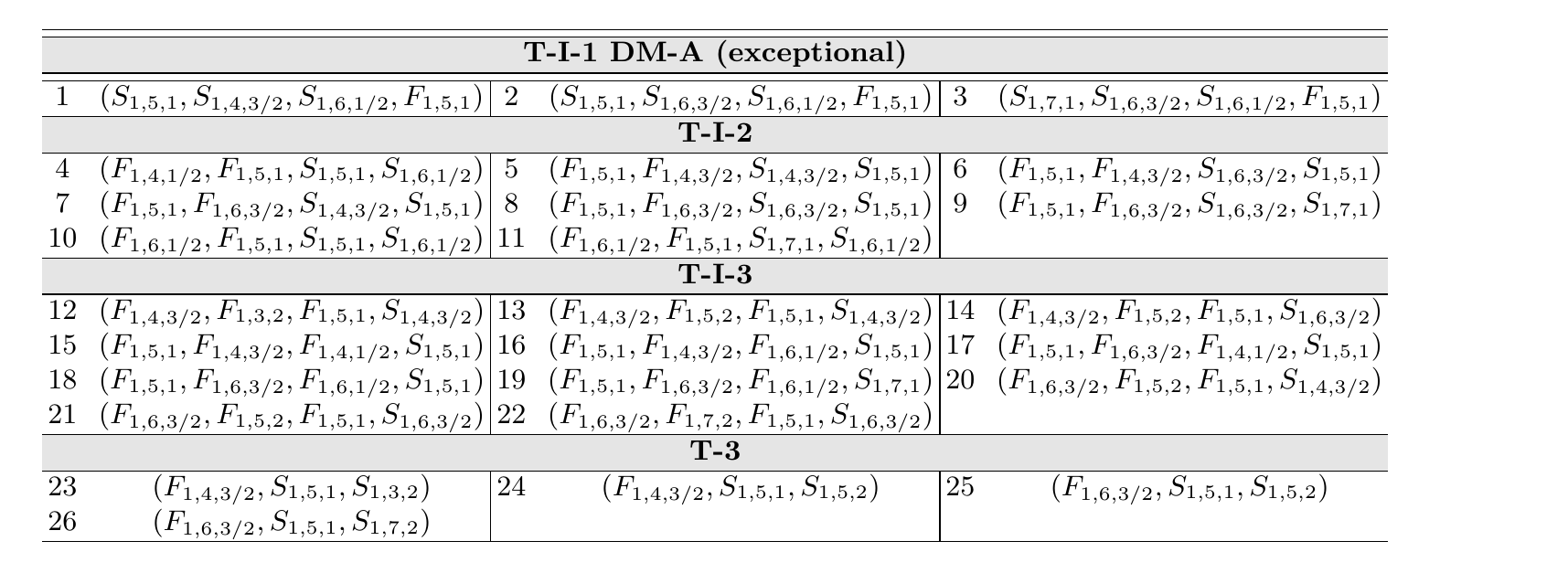}
\caption{Exceptional DM candidates with $Y = 1$. The table at the top corresponds to  DM models with exits (DM-E) which need a stabilizing symmetry and the table at the bottom to accidental DM models (DM-A). }
\label{tab:TableDM-Ex}
\end{table}

\bigskip

\centerline{\bf Acknowledgements}
\medskip
This work is supported by the Spanish grants PID2020-113775GB-I00
(AEI/10.13039/ 501100011033) and CIPROM/2021/054 (Generalitat
Valenciana).  J.C.H.  acknowledge support from grant ANID FONDECYT-Chile
No. 1201673. S.K. is supported by ANID PIA/APOYO AFB180002 (Chile) and by ANID FONDECYT (Chile) No. 1190845. J.C.H. and S.K. acknowledge support from ANID – Programa Milenio – code ICN2019\_044. R.C. is supported by the Alexander von Humboldt Foundation Fellowship. C.A. is supported by FONDECYT-Chile grant
No. 11180722 and ANID-Chile PIA/APOYO AFB 180002.

% Bibliography
\bibliographystyle{BibFiles/utphys}
\bibliography{BibFiles/MyBibTexDatabase}

\end{document}